\pdfoutput=1

\documentclass[11pt]{article}

\usepackage{amssymb,amsmath,amsthm}	
\usepackage{a4wide}
\usepackage{graphicx}      										
\usepackage{tikz}          										
\usepackage{url}
\usepackage{dsfont}
\usepackage{amsmath}
\usepackage{xspace}
\usepackage{subcaption}
\usepackage{doi}
\usepackage[pagewise]{lineno}
\usepackage{cite}

\newcommand{\revised}[1]{\textcolor{black}{#1}}

\newtheorem{remark}{Remark}[]

\newcommand{\R}{\ensuremath{\mathbb{R}}}

\DeclareMathOperator*{\argmin}{argmin}

\begin{document}
	
	\title{Portfolio Optimization and Model Predictive Control:\\ A Kinetic Approach}
	\author{Torsten Trimborn\footnote{Institut f\"ur Geometrie und Praktische Mathematik, RWTH Aachen, Templergraben 55, 52056 Aachen, Germany} \footnote{Corresponding author: trimborn@igpm.rwth-aachen.de}, Lorenzo Pareschi\footnote{Department of Mathematics and Computer Science, University of Ferrara, Via Machiavelli 30, I-44121 Ferrara, Italy}, Martin Frank\footnote{Karlsruhe Institute of Technology, Steinbuch Center for Computing, Hermann-von-Helmholtz-Platz 1, 76344 Eggenstein-Leopoldshafen, Germany}}

	\maketitle


\begin{abstract}
In this paper, we introduce a large system of interacting financial agents in which all agents are faced with the decision of how to allocate their capital between a risky stock or a risk-less bond. The investment decision of investors, derived through an optimization, drives the stock price. The model has been inspired by the econophysical Levy-Levy-Solomon model \cite{levy1994microscopic}. The goal of this work is to gain insights into the stock price and wealth distribution. We especially want to discover the causes for the appearance of power-laws in financial data. We follow a kinetic approach similar to \cite{maldarella2012kinetic} and derive the mean field limit of the microscopic agent dynamics.
The novelty in our approach is that the financial agents apply model predictive control (MPC) to approximate and solve 
the optimization of their utility function. Interestingly, the MPC approach gives a mathematical connection between the two opposing economic concepts of modeling financial agents to be rational or boundedly rational. Furthermore, this is to our knowledge the first kinetic portfolio model which considers a wealth and stock price distribution simultaneously. Due to the kinetic approach, we can study the  wealth and price distribution on a mesoscopic level. The wealth distribution is characterized by a log-normal law. For the stock price distribution, we can either observe a log-normal behavior in the case of long-term investors or a power-law in the case of high-frequency trader. Furthermore, the stock return data exhibit a fat-tail, which is a well known characteristic of real financial data. 
\end{abstract}	

\section{Introduction }	
	
	
The question of allocating capital between a risky and risk-less asset is a well-known issue for private and institutional investors.
This research question has a long tradition in economics: for example the famous works of Markowitz \cite{markowitz1952portfolio} or Merton \cite{merton1969lifetime}. 
Another research field which has received a lot of attention in the last decade is the modeling of financial markets. Several financial crashes (Black Monday 1987, Dot-com Bubble 2000, Global Financial Crisis 2007) have shown the difficulties encountered by many financial market models in replicating financial data properly \cite{colander2009financial, farmer2009economy}.  

Since the 1970s, econometricians have detected persistent empirical patterns in financial data known as \textit{stylized facts}. Stylized facts are universal statistical properties of financial data which can be observed all over the world \cite{cont2001empirical}. The most prominent examples are: the inequality of wealth and fat-tails in the stock return distribution. Several researchers point out that stylized facts play an important role in the creation of financial crisis \cite{lux2008stochastic, sornette2014physics}. For that reason, the discovery of the origin of stylized facts has become a prospering field of economic research. Up to now, this question could only be answered partially and remains widely open \cite{pagan1996econometrics}. 

Besides other approaches in the financial literature, agent-based financial market models aim to reproduce and therefore understand the origins of stylized facts.  
These models consider a large number of interacting financial agents and share more similarities with particle models in physics than with classical asset-pricing models \cite{lux1999scaling, sornette2014physics}.  These models use tools from statistical physics like Monte Carlo simulations and are part of the new research field econophysics. Major contributions in this field are \cite{levy1994microscopic, lux1999scaling}. 
These complex systems of interacting agents are not only inspired by physical theories but also by behavioral finance. 
Thus, agents are modeled to be boundedly rational in the sense of Simon \cite{simon}. These modern market models are capable of reproducing scaling laws \cite{lux1999scaling}. Although there is a controversial discussion if stylized facts can be generally interpreted as scaling laws, this is one motivation to apply tools from physics onto financial models.
Numerical experiments of agent-based models indicate that psychological misperceptions of investors can be accounted to be one reason for the appearance of stylized facts \cite{cross2005threshold, lux2008stochastic}.

The disadvantage of these particle models is the need to study the complex behavior empirically through computer simulations. In addition, many studies have shown \cite{egenter1999finite, zschischang2001some, kohl1997influence, hellthaler1996influence} that in several agent-based models stylized facts may be caused by finite-size effects of the model and are thus only numerical artifacts. To overcome these problems, it is possible to derive kinetic models based on partial differential equations (PDEs) out of the microscopic particle models, which give us the possibility to study the appearance of stylized facts analytically. There are several examples of such a kinetic approach in the literature \cite{maldarella2012kinetic,chayes2009global, burger2013boltzmann, delitala2014mathematical,che2011kinetic, kanazawa2018derivation, kanazawa2018kinetic, bouchaud2000wealth, cordier2005kinetic, cordier2009mesoscopic, during2008kinetic, matthes2008analysis, bisi2009kinetic, CrossMeanField}. We refer also to \cite{pareschi2013interacting} for a recent survey.

The starting point of our work is an agent-based model of financial agents which aim to optimize their investment decision. They are faced with the decision of how to allocate their capital in a risky stock or a risk-less bond. To determine the investment strategy, agents minimize expected missed revenue of their portfolio where they estimate the future stock return by a convex combination of a fundamental and chartist return estimate. The stock price is driven by the aggregated demand of financial agents. To fix the stock price, we use a relaxation of Walras equilibrium law \cite{walras1898etudes}, utilized in many econophysical models \cite{beja1980dynamic, cross2005threshold, lux1999scaling}. 
The microscopic model is inspired by the famous Levy-Levy-Solomon model \cite{levy1994microscopic}. Further, closely related agent-based models have been studied in \cite{chiarella2007heterogeneous, brock1998heterogeneous}. 

\revised{Thanks to a model predictive control (MPC) approximation we are able to obtain an explicit feedback control of the optimization process and derive the investment decision of agents.} This methodology, often applied in the engineering community \cite{CaBo:04,MayneMichalska1990aa}, has been recently applied to kinetic opinion formation models \cite{albi2014kinetic, albi2014boltzmann} but, to our knowledge, never before to a kinetic financial market model. 

More precisely, we consider a large system of coupled constrained optimization problems and, in order to reduce this system to an explicit set of ordinary differential equations (ODEs), we introduce the game-theoretic concept of Nash equilibria and apply MPC. 
From the perspective of agent modeling, we first consider rational financial agents and derive through the MPC approach boundedly rational agents. 

Mathematically, we perform the mean field limit of the microscopic model to derive a mesoscopic description of the dynamics. This means that we look at the limiting case of infinitely many agents and instead of considering each agent individually we can study the dynamics through probability densities. This limit often provides us with Fokker-Planck type equations which enable us to derive analytic solutions and study the long time behavior of the model. Besides other approaches, we apply the Boltzmann methodology as performed in \cite{maldarella2012kinetic} and well described in \cite{pareschi2013interacting}.




\revised{We can emphasize the main novelties of the present approach in the following:
\begin{itemize}
\item The portfolio dynamic in the model is a consequence of an optimization process and is not postulated a-priori accordingly to some heuristic arguments as in \cite{maldarella2012kinetic} and in most agent based models \cite{pareschi2013interacting}. This provides a natural framework for future generalizations and extensions.
\item To our knowledge, this is the first attempt to translate a portfolio model into a kinetic model which contains both wealth and stock price evolutions. Previous results where limited to the separate study of wealth distributions \cite{cordier2009mesoscopic} or stock price dynamic \cite{maldarella2012kinetic}. We thus are able to analyze wealth and stock price distributions simultaneously and study possible interrelations.
\end{itemize}
}

\revised{In the presentation, we consider three modeling stages. First, we analyze an optimal control problem for a  deterministic portfolio model. The motivation behind this is that the optimization process based on the individual agents' strategies has a deterministic nature and must not influence the stochasticity of the model. In other words, the noise in the model can not be controlled by the agents and therefore must appear after the optimization process to characterize the agents' deviations with respect to their optimal behavior.}
Secondly, we add noise to the investment decision of investors and study the outcome on the mesoscopic level. We show that the distribution of wealth in bonds and stocks can be represented in special cases by log-normal distributions. At the final modeling stage, we introduce a new population of financial brokers, equipped with microscopic stock prices. These stock prices are modeled as stochastic differential equations (SDEs). In the mean field limit of infinitely many brokers, we derive a Fokker-Planck equation which enables us to study the stock price distribution. We distinguish between long-term investors and high-frequency traders. In the case of long-term investors, the stock price distribution is of log-normal type, whereas in the case of high-frequency traders we observe an inverse-gamma distribution which satisfies a power-law for large stock prices \cite{maldarella2012kinetic, bouchaud2000wealth,cordier2005kinetic}. 
In addition, we show numerically that the stock return distributions have a fat-tail. 

The rest of the manuscript is organized as follows. In the next section, we first define the microscopic portfolio model. We then apply the MPC approach to simplify the optimization and derive the investment decision of each financial agent. Then, we derive the mean field limit equation in section 3 and analyze the portfolio distribution. As a next step, in section 4, we extend our model by adding noise to the investment decision and analyze the resulting PDE-ODE system. In section 5, we introduce a population of broker, so that the microscopic stock prices are described by a stochastic process. As it has been done for the previous modeling stages, we perform the mean field limit in order to analyze the stock price distribution. In the last section, we give several numerical examples which confirm our analytical findings. We finish the paper with a short discussion of our results and possible model extensions.

\section{Microscopic Model}
We consider $N$ financial agents equipped with their personal monetary wealth $w_i\geq 0$. We assume non-negative wealth, and thus do not allow debts. Agents have to allocate their wealth between a risky asset (stock) and a risk-free asset (bond). The wealth in the risky asset is denoted by $x_i\geq 0$ and the wealth in the risk-free asset by $y_i\geq 0$. Thus, the wealth of the $i$-th agent at time $t>0$ is given by $w_i(t)=x_i(t)+y_i(t)$.

The time evolution of the risk-free asset is described by a fixed non-negative interest rate $r\geq 0$ and the evolution of the risky asset by the stock return,
\[
\frac{\dot S(t)+D(t)}{S(t)},
\]
where $S(t)$ is the stock price at time $t$ and $D(t)\geq 0$ the dividend. We denote all macroscopic quantities with capital letters.  For now, we assume that the stock price and the dividend are given and that the stock price is a differentiable function of time.
Agents can shift capital between the two assets. We denote the shift from bonds into stocks by $u_i$. Notice, that the investment decision $u_i$ implicitly determines the asset allocation of the capital between both portfolios. Thus, we have the dynamics
\begin{subequations}
\begin{align}
& \dot{x}_i(t) = \frac{\dot S(t)+D(t)}{S(t)}\ x_i(t)+u_i(t)\\
& \dot{y}_i(t) = r\ y_i(t)-u_i(t)\\
\label{MacroStockODE}
& \dot{S}(t)= \kappa\ ED_N(t)\ S(t),
\end{align}
\end{subequations}
where the constant $\kappa>0$ measures the market depth and the investment decisions of agents drive the price through the excess demand
$$
ED_N(t) := \frac{1}{N} \sum\limits_{i=1}^N u_i(t).
$$ 
The excess demand is positive if the investors buy more stocks than they sell. 
The ODE \eqref{MacroStockODE} can be interpreted as a relaxation of the well known equilibrium law, supply equals demand, dating back to the economist Walras \cite{walras1898etudes}.

\paragraph{Investment strategy.}
Next, we describe how agents determine their investment strategy. As in classical economic theory, $u_i$ will be a solution of a risk or cost minimization. 
First, in order to make an investment decision, an agent has to estimate future returns. We take two possible strategies into account, a chartist estimate and a fundamentalist estimate. The estimates need to depend on the current stock price. 

Fundamentalists believe in a fundamental value of the stock price denoted by $s^f>0$ and assume that the stock price will converge in the future to this specific value. The investor therefore estimates the future return of stocks versus the return of bonds as
\begin{align*}
K^f:= U_{\gamma}\left( \omega \frac{s^f-S}{S}\right)-r.
\end{align*}
Here, $U_{\gamma}$ is a value function in the sense of Kahnemann and Tversky \cite{kahneman1979prospect} which depends on the risk tolerance $\gamma$ of an investor. A typical example is $U_\gamma(x) = sgn(x) |x|^{\gamma}$ with $0<\gamma<1$ and sign function $sgn$.
The constant $\omega>0$ measures the expected speed of mean reversion to the fundamental value $s^f$. We want to point out that this stock return estimate is a rate and thus $\omega$ needs to scale with time.

Chartists assume that the future stock return is best approximated by the current or past stock return. They estimate the return rate of stocks over bonds by
\begin{align*}
{K}^c:=U_{\gamma}\left( \frac{\dot{S}/\rho+D}{S} \right)-r.
\end{align*}
The constant $\rho>0$ measures the frequency of exchange rates and $D$ is the nominal dividend \cite{maldarella2012kinetic}. 
Both estimates are aggregated into one estimate of stock return over bond return by a convex combination
$$
K = \chi\  K^f + (1-\chi)\ K^c.
$$
As a result, if $K>0$, the investor believes that stocks will perform better and if $K<0$ that bonds will perform better. 
The weight $\chi$ is determined from an instantaneous comparison between the two investment strategies as modeled in \cite{lux1999scaling}. We let
$$
\chi = W(K^f-K^c),
$$
where $W:\R\to [0,1]$ is a continuous function. If for example, $W(x) = \frac12\left(\text{tanh}(x)+1\right)$, the investor optimistically believes in the higher estimate. \revised{Therefore, in our modeling, agents are not split, as usual, into chartists and fundamentalist, but each agent, individually has a potential chartist/fundamentalist behavior accordingly to the value of the weight $\chi$.}

\paragraph{Objective function.}
Finally, we define the minimization problem that determines agent's actions. 
\revised{We assume that agents minimize a quantity proportional to the expected missed revenues in each portfolio. 
 If stocks are believed to be better ($K >0$), then being invested in bonds is bad, and vice versa. Then, $|K|\ y_i$ for $K>0$ is the expected 
 missed  revenue  of  agent $i$ by  having  invested  in  bonds  but  not in stocks. Equivalently, $|K|\ x_i$ for $K<0$ is the  expected  missed  revenue  of  agent
$i$ by  having  invested  in  stocks  but  not in bonds.  Then we weight the expected missed revenue by the wealth in the corresponding
portfolio and define the running cost by
}

\begin{align*}
\Psi_i := 
\begin{cases}
 |K|\ \frac{x_i^2}{2},&\ K<0,\\
  0,&\ K=0,\\
|K|\ \frac{y_i^2}{2},&\ K>0,
\end{cases}
\end{align*}
which can be conveniently rewritten to $\Psi_i = K\cdot \left( -H(-K)\frac{x_i^2}{2} + H(K)\frac{y_i^2}{2} \right)$, where $H$ is the Heaviside step function, zero at the origin. The weighted missed revenue is larger, the larger the estimated difference between returns $K$. Each agent tries to minimize the running costs 
$$
\int_0^T \left( \frac{\mu}{2}u_i(t)^2+\Psi_i(t)\right) dt.
$$
We consider a finite time interval $[0,T]$ and have added a penalty term on transactions. The penalty term is necessary to convexify the problem but is also reasonable, because it describes transaction costs. The transaction costs are modeled to be quadratic which is an often used assumption in portfolio optimization \cite{bertsimas2008robust, mitchell2013rebalancing}.

Hence, in summary, the optimal control problem for the microscopic model is given by
\begin{subequations} \label{microModelOpt}
\begin{align}
& \dot{x}_i(t) = \frac{\dot S(t)+D(t)}{S(t)}\ x_i(t)+u^*_i(t)\\
& \dot{y}_i(t) = r\ y_i(t)-u^*_i(t)\\
& \dot{S}(t) = \kappa\ ED_N(t)\ S(t)\label{relationStock}\\
&u_i^*:= \argmin\limits_{u_i:[0,T]\to \R}  \int_0^T \left( \frac{\mu}{2}u_i(t)^2+\Psi_i(t)\right) dt. \label{opt}
\end{align}
\end{subequations}
Note that, the dynamics are strongly coupled by the stock price in a non-linear fashion.  Since all investors want to minimize their individual cost function, one needs to solve the optimal control problem in a game-theoretic context. We choose the concept of Nash equilibria which will be explained in detail in the next section. 

\subsection{Model predictive control of the microscopic model}
 In case of many agents, we have a large system of optimization problems \eqref{microModelOpt} which in general is very expensive to solve. For that reason, we approximate the objective functional \eqref{opt} by model predictive control (MPC). 
In the MPC framework, one assumes that investors only optimize on a time interval $[\bar{t},\bar{t}+\Delta t]$ for a small $\Delta t>0$ and fixed $\bar{t}$.
One thus assumes that one can approximate the control $u$ on $[0, T]$ by piecewise constant functions on time intervals of length $\Delta t$. 
We can only expect to observe a suboptimal strategy since we perform an approximation of \eqref{opt}, see for example \cite{CaBo:04,MayneMichalska1990aa}. 

We choose the penalty parameter $\mu$ in the running costs to be proportional to the time step so that $\mu = \nu \Delta t$ for some $\nu$. This can be motivated by checking the units of the variables in the cost functional ($K$ is a rate, thus measured in $1/\text{time}$, $\Psi$ is $\text{wealth}^2/\text{time}$, $u$ $\text{wealth}/\text{time}$). We see that the penalty parameter $\mu$ must be a time unit. Furthermore, we substitute the right-hand side of the stock price equation into the stock return. Thus, the constrained optimization problem reads
\begin{align*}
&u_i^*:=\argmin\limits_{u_i:[\bar{t},\bar{t}+\Delta t]\to \R}\int_{\bar{t}}^{\bar{t}+ \Delta t} \left( \frac{\nu \Delta t}{2} u_i^2(t) + \Psi_i(t)\right) dt,\\
&\dot x_i(t) = \kappa\ ED_N(t)\ x_i(t) + \frac{D(t)}{S(t)}\ x_i(t)+ u_i^*,\quad x_i(\bar{t}) =\bar{x}_i,\\
& \dot y_i(t) = ry_i(t) - u_i^*,\quad y_i(\bar{t})=\bar{y}_i,\\ 
&\dot{S}(t) = \kappa\ ED_N(t)\ S(t),\quad S(\bar{t}) = \bar{S}.
\end{align*}
Here, the quantities with a bar denote the initial conditions of the system of ODEs.
\paragraph{Game theoretic setting.}
We want to solve the MPC problem in a game theoretic setting. All agents are coupled by the stock price respectively excess demand $ED_N$. As pointed out previously, it is impossible that all agents act optimal since all agents play a game against each other. 
Thus, a reasonable equilibrium concept is needed to solve the optimal control problem. We want to search for Nash equilibria. In this setting, each agent assumes that the strategies of the other players are fixed and optimal. Thus, we get $N$ optimization problems which need to be solved simultaneously. Hence, we have a $N$-dimensional Lagrangian $L\in\R^N$.  The i-th entry $L_i$ corresponds to the i-th player and reads:

\begin{align*}
& L_i(x_i,y_i, S, u_i, \lambda_{x_i}, \lambda_{y_i}, \lambda_{S})\\
=&\int\limits_{\bar{t}}^{\bar{t}+\Delta t}\left( \frac{\nu \Delta t}{2} u_i^2(t) + \Psi_i(t)\right) dt+\int\limits_{\bar{t}}^{\bar{t}+\Delta t}\dot{\lambda}_{x_i}\ x_i+ \lambda_{x_i}\ \kappa\ ED_N \ x_i + \lambda_{x_i}\ \frac{D}{S} \ x_i +\lambda_{x_i}\ u_i\ dt -\lambda_{x_i}\ \bar{x}_i\\
&+\int\limits_{\bar{t}}^{\bar{t}+\Delta t}\dot{\lambda}_{y_i}\ y_i+ \lambda_{y_i}\ r\ y_i -\lambda_{y_i}\ u_i\ dt-\lambda_{y_i}\ \bar{y}_i
+\int\limits_{\bar{t}}^{\bar{t}+\Delta t}\dot{\lambda}_{S}\  S+ \lambda_{S}\ \kappa\ ED_N\ S \ dt-\lambda_{S}\ \bar{S},
\end{align*}  
with Lagrange multiplier $\lambda_{x_i}, \lambda_{y_i}, \lambda_{S} $. 
Notice that the quantities $(x^*_j,y^*_j,u^*_j),\ j=1,...,i-1,i+1,...,N $ are assumed to be optimal in the i-th optimization and therefore only enter as parameters in the i-th Lagrangian $L_i$. 
Thus, the optimality conditions are given by
\begin{align*}
&\dot x(t) = \kappa\ ED_N(t)\ x_i(t) + \frac{D(t)}{S(t)}\ x_i+ u_i,\  x_i(\bar{t}) =\bar{x}_i,    \\
& \dot y_i(t) = ry_i(t) - u_i(t),\  y_i(\bar{t})=\bar{y}_i,   \\ 
& \dot{S}(t) = \kappa\ ED_N(t)\ S(t),\ S(\bar{t}) = \bar{S}, \\
&\nu\ \Delta t\ u_i(t) = - \lambda_{x_i}(t)- \lambda_{x_i}(t)\ \frac{\kappa }{N}\ x_i(t)  + \lambda_{y_i}(t)- \frac{\kappa}{N}\ S(t)\ \lambda_{S}(t),\\
& \dot{\lambda}_{x_i}(t) = -\kappa\ ED_N(t)\ \lambda_{x_i}(t)- \frac{D(t)}{S(t)}\ \lambda_{x_i}(t)- \partial_{x_i} \Psi_i(t),\\
& \dot{\lambda}_{y_i}(t) = -r \lambda_{y_i}(t)- \partial_{y_i} \Psi_i(t),\\
& \dot{\lambda}_{S}(t) =  \lambda_{x_i}(t)\frac{D(t)}{S^2(t)} x_i-\kappa\ ED_N(t)\ \lambda_{S}(t)- \partial_{S} \Psi_i(t).\\
\end{align*}
We apply a backward Euler discretization to the adjoint equations 
and assume $\lambda_{x_i}(\bar{t}+\Delta t)= \lambda_{y_i}(\bar{t}+\Delta t)= \lambda_{S}(\bar{t}+\Delta t)=0$. This gives
\begin{align*}
&\lambda_{x_i}(\bar{t}) = \Delta t\ \partial_{x_i} \Psi_i(\bar{t}+\Delta t),\\
&\lambda_{y_i}(\bar{t}) = \Delta t\ \partial_{y_i} \Psi_i(\bar{t}+\Delta t),\\
&\lambda_{S}(\bar{t}) = \Delta t\ \partial_{S} \Psi_i(\bar{t}+\Delta t). 
\end{align*}
Hence, the optimal strategy is given by
\begin{align*}
u_N^*(x_i,y_i, S)= 
\begin{cases}
\frac{1}{\nu} ( {K}\ y_i - \frac{\kappa}{N}\ S\  (\partial_{S} {K})\ \frac{y_i^2}{2}),&\quad K>0,\\
0,&\quad K=0,\\
\frac{1}{\nu} ( {K}\ x_i + {K}\ \frac{\kappa}{N}\ x_i^2 + \frac{\kappa}{N} S\  (\partial_{S} {K})\ \frac{x_i^2}{2}),&\quad K<0. 
\end{cases}
\end{align*}
\paragraph{Feedback controlled model.}
The feedback controlled model finally reads
\begin{subequations}
\label{eq:wealthN}
\begin{align}
& \dot{x}_i(t) =\kappa\ ED_N(t)\ x_i(t)+ \frac{D(t)}{S(t)}\ x_i(t)+u_N^*(t,x_i,y_i,S)\\
& \dot{y}_i(t) = r\ y_i(t)-u_N^*(t,x_i,y_i, S) \\
& \dot{S}(t) = \kappa\ ED_N(t)\ S(t).
\end{align}
\end{subequations}
Here, we have inserted the right-hand side of the stock equation \eqref{MacroStockODE} into the stock return.


\section{Mean field limit of feedback controlled model}
In this section, we want to perform the limit of infinitely many agents $N\to\infty$, known as mean field limit. Classical literature on this topic are \cite{braun1977vlasov, dobrushin1979vlasov, neunzert1977vlasov}.
The goal is to derive a mesoscopic description of the financial agents instead of considering each agent in the $N$ particle phase space individually. Thus, instead of considering agents' dynamics in a large dynamical system, we want to describe the dynamics with the help of a density function $f(t,x,y),\ x,y \geq 0$. The density $f(t,x,y)$ describes the probability that an agent at time $t$ has an amount $x\geq 0$ of wealth invested in his or her risky portfolio and  $y\geq 0$ wealth in his or her risk-free portfolio. 

We use the empirical measure
$$
f^N_{(\boldsymbol{x,y})}(x,y):= \frac1N\sum\limits_{k=1}^N \delta(x-x_k)\ \delta(y-y_k),
$$
for given vectors $\boldsymbol{x}:=(x_1,...,x_N)^T\in \R^N$ and $\boldsymbol{y}:=(y_1,...,y_N)^T\in\R^N$
to derive the mean field limit equation formally. We assume that the microscopic model has a unique solution. Furthermore, we denote the solution of the wealth evolution by $\boldsymbol{x}(t):=(x_1(t),...,x_N(t))^T\in \R^N$ and  $\boldsymbol{y}(t):=(y_1(t),...,y_N(t))^T\in \R^N$. We consider a test function $\phi(x,y),\ x,y\geq 0$ and compute
\begin{align*}
\frac{d}{dt} \langle f^N_{(\boldsymbol{x(t),y(t)})} & (t,x,y), \phi(x,y) \rangle = \frac1N \sum\limits_{k=1}^N \frac{d}{dt} \phi(x_k(t),y_k(t))\\
&=\frac1N \sum\limits_{k=1}^N \partial_x\phi(x_k(t),y_k(t))\ \dot{x}_k(t)+ \partial_y\phi(x_k(t),y_k(t))\ \dot{y}_k(t)\\
&=\frac1N \sum\limits_{k=1}^N \partial_x\phi(x_k(t),y_k(t))\  \left( \kappa\ ED_N(t)\ x_k(t)+ \frac{D(t)}{S(t)}\ x_k(t)+u^*(t,x_k,y_k,S) \right)\\
&\quad + \frac1N \sum\limits_{k=1}^N \partial_y\phi(x_k(t),y_k(t))\  \left(r\ y_k(t) -u^*(t,x_k,y_k,S) \right)\\
&=\left\langle f^N_{(\boldsymbol{x(t),y(t)})}(t,x,y),\ \partial_x\phi(x,y)\  \left( \kappa\ ED(t,f,S)\ x+ \frac{D(t)}{S(t)}\ x+u^*(t,x,y,S) \right) \right\rangle \\
&\quad + \left\langle f^N_{(\boldsymbol{x(t),y(t)})}(t,x,y),\ \partial_y\phi(x,y)\ \big(r\ y -u^*(t,x,y,S)\big) \right\rangle. 
\end{align*} 
Here, $\langle \cdot \rangle$ denotes the integration over $x$ and $y$. The excess demand $ED$ and optimal control $u^*$ are given by:
\begin{align*}
ED(t,f^N_{(\boldsymbol{x(t),y(t)})}, S) &:=  \frac{1}{N} \sum\limits_{k=1}^N u^*(t,x_k,y_k, S) = \int \int u^*(t,x,y, S)\ f^N_{(\boldsymbol{x(t),y(t)})}(t,x,y)\ dxdy\\
u^*(t,x,y,S)&:=
\begin{cases}
 \frac{1}{\nu}\  K(t,S)\ x,\ & K<0,\\
  0,\ & K=0,\\
\frac{1}{\nu}\  K(t,S) \ y ,\ & K>0.
\end{cases}
\end{align*}

Hence, the empirical measure $f_{(\boldsymbol{x(t),y(t)})}^N(t,x,y)$ satisfies the equation  
\begin{equation}\label{MF:Feedback}
\begin{aligned}
\partial_t f(t,x,y) + &\partial_x \left( \left[ \kappa\ ED(t,f,S)\ x+ \frac{D(t)}{S(t)}\ x+u^*(t,x,y,S)  \right]   f(t,x,y)\right)\\
+& \partial_y\left( [r\ y+ u^*(t,x,y,S)] f(t,x,y) \right) =0,
\end{aligned}
\end{equation}
in the weak sense. 
We call the PDE \eqref{MF:Feedback} the \textbf{mean field portfolio equation}.
Thus the mean field portfolio stock price evolution is described by the PDE \eqref{MF:Feedback} coupled with the macroscopic stock price ODE
$$
\dot{S}(t) = \kappa\ ED(t,f,S)\ S(t). 
$$



The mesoscopic behavior can be studied by the mean field portfolio equation. Compared to models, which only consider ODEs, this is a huge benefit of the kinetic approach. 
We define the following marginals of $f$:
$$
g(t,x):= \int f(t,x,y)\ dy,\quad h(t,y):=\int f(t,x,y)\ dx.
$$
The corresponding moments read: (2.2)
$$
X(t):= \int x\ g(t,x)\  dx,\quad Y(t):= \int y\ h(t,y)\  dy.
$$
Hence, $g$ is a probability density function of the wealth invested in stocks and 
$h$ is the probability density function of wealth invested in bonds.
We then integrate the mean field portfolio equation over $y$ respectively $x$ to observe equations for $g$ and $h$. Since the optimal control $u^* $ depends on both microscopic quantities, we cannot expect to get a closed equation for $g$ or $h$ in general. 
Nevertheless, in the special case $K<0$, the control $u^*$ only depends on $x$ and the time evolution of $g$ reads
\begin{align*}
\partial_t g(t,x) &+ \partial_x\left( \left[ \frac{K(S(t))}{\nu} \left(\kappa\ X(t) +1\right) + \frac{D(t)}{S(t)}\right] x\ g(t,x)\right)  = 0.
\end{align*}
One solution of the equation is given by:
$$
g(t,x)= \frac{{c}}{\sqrt{\pi} x}\ \exp\left\{-\left(\log(x)-\int\limits_0^t \frac{K(S(\tau))}{\nu} \left(\kappa\ X(\tau) +1\right) + \frac{D(\tau)}{S(\tau)} \ d\tau\right)^2\right\},\ {c}>0.
$$
Notice that $g$ is the distribution function of a log-normal law.
We get a closed equation for $h$, in the case $K>0$, in the same way. The solution $h$ is of log-normal type as well. 
We refer to the Appendix \ref{appMar} for a detailed discussion. 

\revised{\begin{remark}
Clearly, even if the agents are indistinguishable in the mean field limit, further individual features can be potentially inserted in their microscopic dynamic. For example, the agents' attitude towards risk or other behavioral aspects. In the large number of agents limit this will lead to the presence of further independent variables in the statistical distribution of agents $f$. 
\end{remark}
}

\section{Feedback controlled model with noise}
\revised{So far, we have considered a fully deterministic setting. This choice has been dictated by our modeling strategy based on solving an optimal control problem for the agents' portfolio. The possible stochasticity of the model, in fact, should be kept out of the control capabilities of the agents optimization process. On the other end, in a realistic stock market model, the presence of randomness is an essential feature. It is generally accepted that stock prices are unpredictable and e.g.\ news and political decisions influence the behavior of market participants in an uncertain fashion. For that reason, we discuss the effect of randomness to the feedback controlled microscopic model \eqref{eq:wealthN} and to its mean field limit.}
 
The optimal control of the i-th agent was given by
$$
u^{\star}(x_i,y_i,S) =
\begin{cases}
 \frac{1}{\nu}\  K(S) \ x_i ,&\ K<0,\\
 0,&\ K=0,\\
\frac{1}{\nu}\  K(S) \ y_i ,&\ K>0.
\end{cases}
$$
Notice that the investment decision of agents only differs through different personal wealth. Thus, the estimate of stock return over bond return was identical for all investors. This assumption seems to be too simple, so each individual should differ in their return estimate. Hence, we add white noise to the returns estimate. Since the return estimate is a rate, the random variable also needs to scale with time. We use symbolic notation of integrals adopted from the common notation of SDEs to define the integrated noisy optimal control 

$$
u^{\star}_{\eta_i}(x_i,y_i,S)\ dt =
\begin{cases}
\frac{1}{\nu} K(S) \ x_i\ dt+ \frac{1}{\nu}  x_i\   dW_i\,&\ K<0,\\
 0,&\ K=0,\\
\frac{1}{\nu}   K(S)\ y_i \ dt+\frac{1}{\nu}\ y_i\  dW_i  ,&\ K>0.
\end{cases}
$$


Here, $dW_i$ denotes the stochastic It$\hat{\text{o}}$ integral and thus the feedback controlled microscopic system with noise is given by 
\begin{subequations} \label{microMC}
\begin{align}
& d{x}_i = \left( \kappa ED_N \ x_i+ \frac{D}{S}\ x_i+u^*_i \right)\ dt+ \frac{1}{\nu} ( H(-K) x_i+ H(K) y_i)\ dW_i\\
& d{y}_i = \left( r\ y_i-u^*_i\right)\ dt - \frac{1}{\nu} ( H(-K) x_i+ H(K) y_i)\ dW_i\\
&d{S} = \left( \kappa ED_N\ S\right)\ dt.
\end{align}
\end{subequations}

\subsection{Mean field limit}
The goal of this section is to derive a mesoscopic description of the particle dynamics with noise. The classical mean field approaches by Braun, Hepp, Neunzert and Dobrushin \cite{neunzert1977vlasov, braun1977vlasov, dobrushin1979vlasov} do not apply because of the white noise. The only known mean field result in the case of diffusion processes is the convergence of $N$ interacting processes to the kinetic McKean-Vlasov equation \cite{sznitman1991topics}. 
Unfortunately, the model \eqref{microMC} does not satisfy the classical assumptions since the $N$-particle dynamics are coupled with the macroscopic stock price ODE. 

\begin{figure}[ht]
\begin{center}
\includegraphics[scale=0.8]{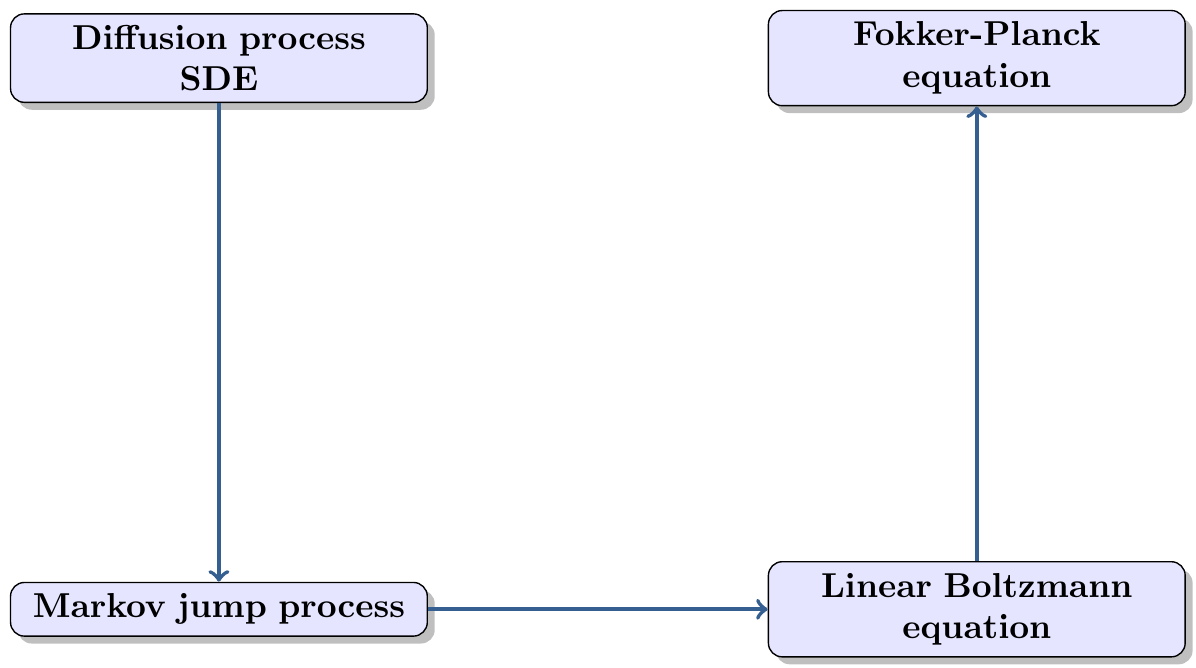} 
\end{center}
\caption{Sketch of the modelling process.}
\label{diagram}
\end{figure}

The following modeling approach, well described in \cite{pareschi2013interacting}, is an alternative method to derive the mean field limit of the microscopic model \eqref{microMC} at least formally. The idea is to discretize the diffusion process and interpret it as a Markov jump process.
Then, one can derive the corresponding master equation, which can be also interpreted as a linear Boltzmann equation. With the right scaling, known in kinetic theory as grazing limit, one observes in the limit the Fokker-Planck equation (See Figure \ref{diagram}).

\paragraph{Boltzmann model.}
As seen before, we consider the probability density $f(t,x,y)$ which describes an investor to have monetary wealth $x\geq 0$ in his or her risky portfolio and wealth  $y\geq 0$ in his or her  risk-free portfolio. 
The portfolio dynamics are characterized by the following linear interactions $(x,y)\mapsto (x^{\prime},y^{\prime})$.
\begin{align*}
&x^{\prime} = x+a\ \left( \kappa\ ED(t)+\frac{D(t)}{S(t)}\right) x + a\ u_{\eta}^*(t,x,y,S),\ &&\text{if}\ x^{\prime}>0,\\
&y^{\prime} = y+ a\ y\ r- a\ u_{\eta}^*(t,x,y,S),\ &&\text{if}\ y^{\prime}>0,
\end{align*} 
with
\begin{align*}
& u^*_{\eta}:= u^*(t,x,y,S)+ \frac{1}{\sqrt{a}\ \nu}\ (H(-K)\ x +H(K)\ y )\ \eta,\\
& a:= \Delta t, 
\end{align*}
and a normally distributed random variable $\eta$ with zero mean and variance one. The time step $\Delta t>0$ is fixed and originates from the Euler-Maruyama discretization  of the SDE. The time evolution of the density function $f(t,x,y)$ is then described by an integro-differential equation of Boltzmann type. In weak form, the equation reads
\begin{align}
&\frac{d}{dt}\ \int \phi(x,y)\ f(t,x,y)\ dxdy = (L(f),\phi), \label{boltzmann}\\
& (L(f),\phi):= \left\langle \int \mathcal{K}(x,y,S,ED,D,\eta)\ (\phi(x^{\prime},y^{\prime})-\phi(x,y))\ f(t,x,y)\  dx dy \right\rangle,
\end{align}
with a suitable test function $\phi(x,y)$ and $\langle \cdot \rangle$ denotes the expectation with respect to the random variable $\eta\in \R$. The interaction kernel $\mathcal{K}$ has to ensure that the post interaction portfolio values remain positive:
\[
\mathcal{K}(x,x^{\prime},y,y^{\prime},S,ED,D,\eta):= \theta\ \mathds{1}_{\{x^{\prime}> 0 \}} \mathds{1}_{\{y^{\prime}> 0 \}}\ \eta,
\]
where $\theta>0$ is the collision rate and $\mathds{1}(\cdot)$ the indicator function. The interaction kernel can be simplified if there is no dependence on $x$ and $y$.
This case corresponds to the case of Maxwellian molecules in the classical Boltzmann equation. This can be achieved by truncating the random variable $\eta$ in a way that the post interaction wealth always remains positive.
In our case, it is not possible to state explicit bounds for the random variable $\eta$ since the stock return is not bounded. In fact, for a sufficiently small step size $\Delta t$, it is always possible to truncate the random variable in a way that the kernel is independent of $x,y$. Then, the interaction operator reads:
\[
(L(f),\phi):= \left\langle \theta \int \ (\phi(x^{\prime},y^{\prime})-\phi(x,y))\ f(t,x,y)\  dx dy \right\rangle.
\]
We can immediately observe that the model conserves the number of agents, which corresponds to the choice $\phi(x,y)=1$. We are interested in the asymptotic behavior of the density function $f$. 
\paragraph{Asymptotic limit.}
The goal of the asymptotic procedure is to derive a simplified model of Fokker-Planck type. Thus, the integral operator gets translated into a second order differential operator. The procedure can be described in two steps. First, we perform a second order Taylor expansion of the test function $\phi(x^{\prime},y^{\prime})$. Secondly, we rescale characteristic parameters of the model, preserving the main macroscopic properties of the original kinetic equation \eqref{boltzmann}. A closely related approach in kinetic theory is the grazing collision limit \cite{villani1998new}. We introduce the scaling 
$$
\theta = \frac{1}{\epsilon},\quad a= \epsilon,
$$
where $\epsilon>0$ and perform the limit $\epsilon\to 0$. The limit equation is given by the the following Fokker-Planck equation
\begin{align*}
&\partial_t f(t,x,y)+\partial_x((\kappa\  ED(t,f,S)\ x + \frac{D(t)}{S(t)}\  x+u^*(t,x,y,S))\ f(t,x,y) )\\
&+\partial_y((r\ y-u^*(t,x,y,S))\ f(t,x,y))+\frac{1}{2\ \nu^2}\partial_{yx}^2\Big( (H(-K)x+ H(K)y)^2\ f(t,x,y)\Big) \\
&=\frac{1}{2\ \nu^2} \partial_x^2\Big( (H(-K)x+ H(K)y)^2\ f(t,x,y)\Big)+\frac{1}{2\ \nu^2}\partial_y^2\Big( (H(-K)x+ H(K)y)^2\ f(t,x,y)\Big) ,
\end{align*}
coupled with the macroscopic stock price ODE
$$
\dot{S}(t) = \kappa\ ED(t,f,S)\ S(t).
$$
We call the previously introduced PDE the \textbf{diffusive mean field portfolio equation}.
In the Appendix \ref{asymp}, we provide a detailed derivation.

\subsection{Marginals of diffusive mean field portfolio equation}\label{MargDiff}
Again, we are interested in the behavior of the marginal distributions $g$ and $h$. 
In the special case $K<0$, the control $u^*$ only depends on $x$ and the time evolution of $g$ reads.
\begin{align*}
\partial_t g(t,x) &+ \partial_x\left( \left[\frac{K(S(t))}{\nu} \left(\kappa\ X(t) +1\right)+\frac{D(t)}{S(t)}\right] x\ g(t,x)\right) - \partial_x^2\left( \frac{x^2}{2\ \nu^2} g(t,x) \right)  = 0.
\end{align*}
In order to search for self-similar solutions, we introduce the scaling $\bar{g}(t,\bar{x}) = x\ g(t,x),\ \bar{x}=\log(x)$ and define $b(t):= \frac{K(S(t))}{\nu} \left(\kappa\ X(t) +1\right)+\frac{D(t)}{S(t)} $. We observe a linear convection-diffusion equation for the evolution of $\bar{g}(t,\bar{x})$
\begin{align*}
\partial_t \bar{g}  (t,\bar{x})+ \left(b(t)-\frac{1}{2\ \nu^2}\right) \ \partial_{\bar{x}} g(t,\bar{x}) = \frac{1}{2\ \nu^2} \partial_{\bar{x}}^2 \bar{g}(t,\bar{x}).
\end{align*}
The solution is given by
\begin{align*}
\bar{g}(t,\bar{x})= \frac{1}{(2\ \left(\frac{t}{\nu^2}+c \right)\ \pi)^{\frac12}} \exp\left\{- \frac{(\bar{x}+\frac{\frac{t}{\nu^2}+c}{2}-B(t))^2}{2\ (\frac{t}{\nu^2}+c)}\right\},\ c>0,
\end{align*}
with $B(t):=\int\limits_0^t b(\tau)\ d\tau +\bar{c},\ \bar{c}>0$. After reverting to the original variables, we get
$$
g(t,x)= \frac{1}{x\ (2\ (\frac{t}{\nu^2}+c)\ \pi)^{\frac12}}  \exp\left\{ -\frac{\left(\log(x)+\frac{\frac{t}{\nu^2}+c}{2}-B(t)\right)^2}{2\ (\frac{t}{\nu^2}+c)}\right\},\ c>0.
$$
Thus, the wealth in bonds admits a log-normal asymptotic behavior as well. \\
Analogously, we obtain a similar equation for $h$ in the case $K>0$. The solution also satisfies a log-normal law. 
For details, we refer to the Appendix \ref{MarDiff}. At first glance, we did not gain any new information compared to the marginals of the mean field portfolio equation. In both cases, we have observed log-normal behavior. However, this is not true, in the diffusive case, the solution admits a time dependent variance and is not constant in contrast to the deterministic case. In addition, we have observed that adding multiplicative noise does not change the portfolio distribution drastically. 

\section{Stock price as random process}
Until now, the macroscopic stock price evolution has been given by the ODE \eqref{MacroStockODE} and was deterministic. 
We aim to analyze the price behavior in a probabilistic setting and analyze the price distribution. 
We modify the model by adding a microscopic stochastic model beneath the macroscopic stock price equation \eqref{MacroStockODE}. To do so,  we introduce a new population of $M$ brokers or reference traders. Each broker is equipped with a microscopic stock price $s_j>0$. The microscopic stock prices are modeled as random processes. The average of broker prices generates the macroscopic stock price 
$$
S_M:=\frac{1}{M} \sum\limits_{j=1}^M s_j.
$$
 The stochastic nature of microscopic stock prices can be explained by different market accessibility of each broker. Their individual stock price is given by
\begin{align}\label{brokersys}
d{s}_j= \kappa\ ED\ s_j\ dt+ s_j\ dW_j,\quad j=1,...,M,
\end{align} 
where $W_j$ is a Wiener process and  equation \eqref{brokersys} has to be interpreted in the It$\hat{\text{o}}$ sense. Compared to the macroscopic stock price equation \eqref{MacroStockODE}, there is multiplicative noise added to the price evolution of brokers.

Now, the stock price evolution is coupled with the portfolio evolution in two different ways. First, by the stock return in the stock portfolio and secondly by the investment decision $u^*$ 
\begin{align*}
&\partial_t f+\partial_x((\kappa\  ED\ x + \frac{D}{S}\  x+u^*)\ f )+\partial_y((r\ y-u^*)\ f)+\frac{1}{2\ \nu^2}\partial_{yx}^2\Big( (-H(-K)x+ H(K)y)^2\ f\Big) \\
&\quad\quad =\frac{1}{2\ \nu^2} \partial_x^2\Big( (H(-K)x+ H(K)y)^2\ f\Big)+\frac{1}{2\ \nu^2}\partial_y^2\Big( (H(-K)x+ H(K)y)^2\ f \Big) ,\\
& d{s}_j= \kappa\ ED\ s_j\ dt+ s_j\ dW_j,\quad j=1,...,M.
\end{align*} 
We need to specify whether the investors' decisions are based on the microscopic or macroscopic stock price. The macroscopic stock price determines the stock return of agents' portfolios, because this is the global market price.
In the case of the investment decision, one can argue that an investor might trade on the microscopic or macroscopic stock price. Arbitrage opportunities are a reason to act on the microscopic scale. In addition, one can argue that the microscopic stock prices have in fact a smaller time scale than the macroscopic stock price since the latter is the average of the former. This leads us to the characterization that investors acting on the micro prices are \textbf{high-frequency traders}, whereas agents acting on the macro price can be accounted to be \textbf{long-term investors}. 
\paragraph{Mean field limit.}
As seen before, we want to consider the mean field limit of the microscopic stock price equations. In fact, the microscopic brokers only differ in their initial conditions and multiplicative noise. We have:
\begin{align}
d{s}_j(t)= \kappa\ ED(t,f,\boldsymbol{s})\ s_j(t)\ dt+ s_j(t)\ dW_j,\quad s_j(0) = s_j^0.\label{brokerSDE}
\end{align} 
 Thus, there is no coupling between brokers and we have a simple setting of McKean-Vlasov type equations. We have written the excess demand as $ED(t,f,\boldsymbol{s})$
since we can have $ED(t,f,s_j)$ in the high-frequency case or $ ED(t,f,S_M)$ for long-term investors. We assume that the empirical measure $V^N_{\textbf{s(0)}}(0,s)$, which is defined by the initial conditions of the microscopic system 
 \[
 V^N(0,s) := \frac{1}{M}\sum\limits_{k=1}^M \delta(s-s_k^0),
 \]
converges to a distribution function $V(0,s)$. Then, the system \eqref{brokersys} converges in expectation to the mean field SDE
\begin{align}
d{\bar{s}}_j(t)= \kappa\ ED(t,f,\bar{s}_j)\ \bar{s}_j(t)\ dt+ \bar{s}_j(t)\ dW_j,\quad \bar{s}_j(0) = \mathfrak{s}_j,\quad \mathfrak{s}_j\sim V(0,s).\label{MFSDE}
\end{align} 
Due to the Wiener process, the above set of stochastic processes is independent and in particular identically distributed. We can thus apply the Feynman-Kac formula and the distribution $V(t,s),\ s>0$ evolves accordingly to 
\begin{align} \label{stocklimit}
\partial_t V(t,s) + \partial_s\left( \kappa\ ED(t,f,(\cdot))\ s\ V(t,s)\right) = \frac{1}{2}\  \partial_s^2(s^2\ V(t,s)). 
\end{align}
Notice that the macroscopic stock price is the first moment of $V$.
$$
S(t)=\int s\ V(t,s)\ ds. 
$$
Hence, the \textbf{diffusive mean field portfolio stock price} system is given by:
\begin{align*}
&\partial_t f(t,x,y)+\partial_x((\kappa\  ED(t,f,S)\ x + \frac{D(t)}{S(t)}\  x+u^*(t,x,y, (\cdot)))\ f(t,x,y,s) )\\
&\quad\quad  -\frac{1}{2\ \nu^2} \partial_x^2\Big( (H(-K)x+ H(K)y)^2\ f(t,x,y)\Big)   + \frac{1}{2\ \nu^2}\partial_{yx}^2\Big( (H(-K)x+ H(K)y)^2\ f(t,x,y)\Big)\\
&\quad\quad +\partial_y((r\ y-u^*(t,x,y, (\cdot)))\ f(t,x,y)) - \frac{1}{2\ \nu^2}\ \partial_y^2\Big( (H(-K)x+ H(K)y)^2\ f(t,x,y)\Big)=0,\\
&\partial_t V(t,s) + \partial_s\left( \kappa\ ED(t,f,(\cdot))\ s\ V(t,s)\right) = \frac12\  \partial_s^2(s^2\ V(t,s)). 
\end{align*}
Remember that the influence of the investment decision enters in the stock-price evolution through the excess demand. In the next sections, we want to study the influence of a high-frequency or long-term strategy of investors on the price distribution $V(t,s)$.

\subsection{Long-term investors}
In the case of long-term investors, the investment decision $u^*=u^*(t,x,y,S)$ depends on the macroscopic stock price $S$. The stock price equation is given by:
\begin{small}
\begin{align}\label{stockPDElong}
\partial_t V + &\partial_s\left( \frac{\kappa}{\nu}\ K(S)\ \left[\int \int [H(-K(S))x+H(K(S))y]\ f(t,x,y)\ dxdy\right]\ s\ V\right) = \frac{1}{2}\  \partial_s^2(s^2\ V). 
\end{align}
\end{small}
We can take the first moment of the previous equation and obtain the macroscopic stock price ODE \eqref{MacroStockODE} considered in the previous sections.

\paragraph{Asymptotic behavior.} Due to the fact that the stock price is a stochastic process, we can study the distribution function of the stock price PDE. We define
\begin{align*}
&P(t):=\int \int [H(-K(S))x+H(K(S))y]\ f(t,x,y)\ dxdy,\\
& R(t):= \frac{\kappa}{\nu}\ K(S(t))\ P(t),  
\end{align*}
and search for self-similar solutions of equation \eqref{stockPDElong}. The quantity $P$ is the average amount of wealth in the bond or stock portfolio and $R$  is the average amount of wealth invested in stocks. We consider the scaling $ \mathcal{V}(p,t) = s\ V(t,s),\ p=\log(s)$ and $\mathcal{V}$ thus satisfies the following linear convection-diffusion equation
\begin{align*}
\partial_t \mathcal{V}(t,p) + \left(  R(t)-\frac12  \right)\ \partial_p \mathcal{V}(t,p) = \frac12 \partial^2_p \mathcal{V}(t,p).  
\end{align*} 
The solution of the previous equation is given by
$$
\mathcal{V}(t,p) = \frac{1}{\sqrt{2\ \pi}} \exp\left\{ -\frac{(p+\frac{t+c_1}{2} - \bar{R}(t))^2}{2}\right\},
$$
for a constant $c_1>0$ and $\bar{R}(t):=\int\limits_0^t R(\tau)\ d\tau +c_2,\ c_2>0$. 
Hence, by reverting to the original variables, we get
\begin{align*}
V(t,s) = \frac{1}{s\ \sqrt{2\ \pi\ (t+c_1)}} \exp\left\{- \frac{(\log(s)+\frac{t+c_1}{2}-\bar{R}(t))^2}{2\ (t+c_1)} \right\}.
\end{align*}
We thus observe log-normal asymptotic behavior of the model. 
\subsection{High-frequency traders}
In the case of high-frequency investors, we have to clarify the dependence of the optimal control $u^*$ on the microscopic stock price $s$. The investment strategy of \textbf{ high-frequency fundamentalists} can be translated one to one. We have
$$
k^f(t,s):=U_{\gamma}\left( \omega \frac{s^f(t)-s}{s}\right)-r.
$$
The chartist estimated return is more difficult. In fact, the chartists estimate involves a time derivative of the stock price. On the microscopic level, we can insert the right-hand side of the microscopic stock price equation. In addition, we assume that the investor averages over the uncertainty.  Thus, for \textbf{high-frequency chartists} we get:
$$
{k}^c(t,s):=U_{\gamma}\left( \frac{\kappa/\rho \  ED(t,f,s)+D(t)}{s} \right)-r.
$$
We define the aggregated high-frequency estimate of stock return over bond return by
$$
k(t,s):=\chi\ k^f(t,s)\ +  (1-\chi)\ k^c(t,s).
$$
Hence, the \textbf{high-frequency stock price equation} reads
\begin{align*}
\partial_t V(t,s) + \partial_s  \Big( \frac{\kappa}{\nu}\ k(t,s)\  \Big[\int \int [H(-k(t,s))\ x+H(k(t,s))\ y]\  f(t,x,y) & \ dxdy\Big]\ s\ V(t,s)\Big) \\
&= \frac{1}{2}\  \partial_s^2(s^2\ V(t,s)). 
\end{align*}
Notice that we cannot find a closed equation for the first moment of this equation. In general, it is difficult to solve the high-frequency stock price equation. We want to study admissible states of the high-frequency stock price equation in order to obtain a solution. 

In addition, we have to specify the dependence of the diffusive mean field portfolio equation on the microscopic stock price. 
 We want to point out that the diffusive mean field portfolio equation, solely coupled with the high-frequency stock price equation, is not well-defined. This is because of the fact that it is unclear how to interpret the variable $s$ in the optimal control of the diffusive mean field portfolio equation. One solution to this problem is to add the mean field SDE \eqref{MFSDE} to the model. 
The diffusive mean field portfolio equation is then coupled with the mean field SDE through the microscopic stock prices $\bar{s}$ in the optimal control. In addition, the diffusive mean field portfolio equation is coupled with the high-frequency stock price equation by the macroscopic stock price $S$. We get:
\begin{align*}
&\partial_t f(t,x,y)+\partial_x((\kappa\  ED(t,f,S)\ x +u^*(t,x,y, \bar{s}))\ f(t,x,y) )\\
&-\frac{1}{2\ \nu^2} \partial_x^2\Big( (H(-K)x+ H(K)y)^2\ f(t,x,y)\Big)+ \frac{1}{2\ \nu^2}\partial_{yx}^2\Big( (H(-K)x+ H(K)y)^2\ f(t,x,y)\Big)\\
&\quad+\partial_y((r\ y-u^*(t,x,y,\bar{s}))\ f(t,x,y)) - \frac{1}{2\ \nu^2}\ \partial_y^2\Big( (H(-K)x+ H(K)y)^2\ f(t,x,y)\Big)=0,\\
& d\bar{s}(t)= \kappa\ ED(t,f,\bar{s})\ \bar{s}(t)\ dt+ \bar{s}(t)\ dW,\\
 &\partial_t V(t,s) + \partial_s\left( \frac{\kappa}{\nu}\ k(t,s)\ \left[\int \int [H(-k(t,s))x+H(k(t,s))y]\ f(t,x,y)\ dxdy\right]\ s\ V(t,s)\right)\\
 &\quad = \frac{1}{2}\  \partial_s^2(s^2\ V(t,s)).
\end{align*}
Since the solution of the high-frequency stock price equation is the density of the stochastic process $\bar{s}$, we can substitute this PDE by the expected value of the stochastic process $\bar{S}$. The alternative model reads:
\begin{align*}
&\partial_t f(t,x,y)+\partial_x((\kappa\  ED(t,f,S)\ x + \frac{D(t)}{S(t)}\  x+u^*(t,x,y, \bar{s}))\ f(t,x,y) )\\
&-\frac{1}{2\ \nu^2} \partial_x^2\Big( (H(-K)x+ H(K)y)^2\ f(t,x,y)\Big)+ \frac{1}{2\ \nu^2}\partial_{yx}^2\Big( (H(-K)x+ H(K)y)^2\ f(t,x,y)\Big)\\
&\quad+\partial_y((r\ y-u^*(t,x,y,\bar{s}))\ f(t,x,y)) - \frac{1}{2\ \nu^2}\ \partial_y^2\Big( (H(-K)x+ H(K)y)^2\ f(t,x,y)\Big)=0,\\
& d\bar{s}(t)= \kappa\ ED(t,f,\bar{s})\ \bar{s}(t)\ dt+ \bar{s}(t)\ dW,\\
 &S=E[\bar{s}].
\end{align*}
We consider the former model instead of the latter as we can analyze the stock price distribution due to the high-frequency stock price equation. 
 \paragraph{Steady state.} We aim to study the admissible steady states of the high-frequency stock price equation. Under the assumption that a universal stock price distribution exists we are able to characterize the asymptotic behavior. In fact, in special cases the steady state distribution is described by an inverse-gamma distribution. We assume that 
$ \chi\equiv 1,\ s^f\equiv c>0,\ U_{\gamma}(x)=x$ holds. 
The stock price equation is then simplified to
\begin{align*}
\partial_t V(t,s) + &\partial_s\left( \frac{\kappa}{\nu}\ [\omega\ s^f-s\ (\omega+r)]\ \left[\int \int [H(-k(s))x+H(k(s))y]\ f(t,x,y)\ dxdy\right]\ V(t,s)\right) \\
&= \frac{1}{2}\  \partial_s^2(s^2\ V(t,s)).
\end{align*}
Furthermore, we assume that the portfolio distribution $f$ has reached a steady state $f_{\infty}$.
We define 
\begin{align*}
&P_x^{\infty}:= \int \int x\ f_{\infty}(x,y)\ dxdy>0,\quad k<0,\\
&P_y^{\infty}:= \int \int y\ f_{\infty}(x,y)\ dxdy>0,\quad k>0,
\end{align*}
and assume $P^{\infty}:=P_x^{\infty}=P_y^{\infty}$. This assumption implies that the mean wealth in bonds and stocks is constant. Economically, the mean wealth never reaches a steady profile, although it is reasonable that the mean wealth only has minor variations.
Hence, the steady state distribution $V_{\infty}(s)$ satisfies
\begin{align}\label{steadystatefat}
\frac{1}{2} \partial^2_s(s^2\ V_{\infty}(s))- \frac{\kappa}{\nu}\  P^{\infty}\ \partial_s\left([\omega\ s^f-s\ (\omega+r)]\  V_{\infty}(s)\right)=0.
\end{align}
The solution of \eqref{steadystatefat} is given by the inverse-gamma distribution
$$
V_{\infty}(s)= C\ \frac{1}{s^{2(1+\frac{\kappa}{\nu}\ P_{\infty} (\omega+r))}}\ \exp\left\{ -\frac{2\ \frac{\kappa}{\nu}\ \omega\ P_{\infty}\ s^f}{s}    \right\},\ s>0,
$$
where the constant $C$ should be chosen as
$$
C:= \frac{(2\ \kappa\ \omega\ P_{\infty}^x\ s^f )^{1+2 \frac{\kappa}{\nu}\ P_{\infty} (\omega+r)}}{\Gamma(1+2 \frac{\kappa}{\nu}\ P_{\infty} (\omega+r))},
$$
such that the mass of $V_{\infty}$ is equal to one. Here, $\Gamma(\cdot)$ denotes the gamma function. We immediately observe that for large stock prices $s$, the distribution function asymptotically satisfies
$$
V_{\infty} \sim \frac{1}{s^{2(1+\frac{\kappa}{\nu}\ P_{\infty} (\omega+r))}}.
$$
Hence, the equilibrium distribution is described by a power-law. 
\begin{remark}~
\begin{itemize}
\item We have observed that the presence of high-frequency fundamentalists leads to power-law behavior in the stock price distribution.
This coincides with earlier findings in \cite{maldarella2012kinetic}. 
\item The universal features which create power-law tails are multiplicative noise and additionally an external force on the microscopic level. In our case, this force is given by the fundamental value $s^f$ of the fundamental trading strategy.
\end{itemize}
\end{remark}

\section{Numerical Results}
In this section, we present some numerical examples illustrating the behavior of the resulting mean field model. 
We always consider the full kinetic model, namely the diffusive mean field portfolio stock price model.
Our simulations have been conducted with a standard Monte Carlo solver \cite{pareschi2013interacting}. We choose the value function $U_{\gamma}$ and the weight function $W$ as follows:
 \begin{align*}
 &W(K^f-K^c):= \beta\ \left(\frac{1}{2} \tanh\left(\frac{K^f-K^c}{\alpha}\right)+\frac12\right) +(1-\beta)\ \left(\frac{1}{2} \tanh\left(-\frac{K^f-K^c}{\alpha}\right)+\frac12\right),\\
 &\quad\quad  \alpha>0,\  \beta\in[0,1],\\
 &
 U_{\gamma}(x):= \begin{cases}
 x^{\gamma+0.05},\quad  x>0,\\
 -(|x|)^{\gamma-0.05},\quad x\leq 0,\quad \gamma\in [0.05,0.95].
 \end{cases}
 \end{align*}
The weight function $W$ models the instantaneous comparison of the fundamental and chartist return estimate. 
The constant $\beta\in [0,1]$ determines if the investor trusts in the higher ($\beta=1$) or lower estimate ($\beta=0$) and we thus call this constant the trust coefficient. The constant $\alpha>0$ simply scales the estimated returns.
\begin{figure}[ht]
\begin{center}
\includegraphics[scale=0.32]{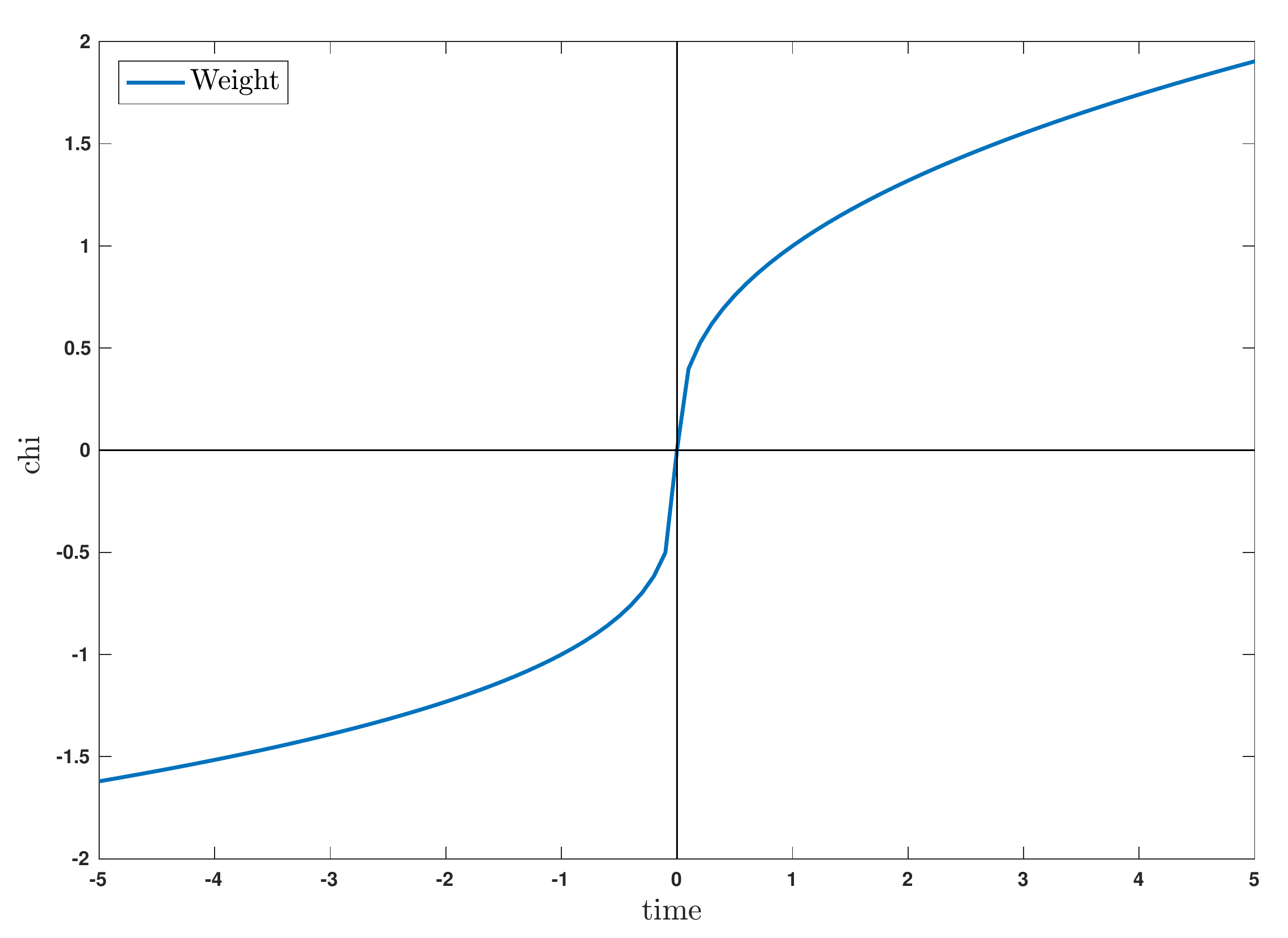}
\hfill
\includegraphics[scale=0.32]{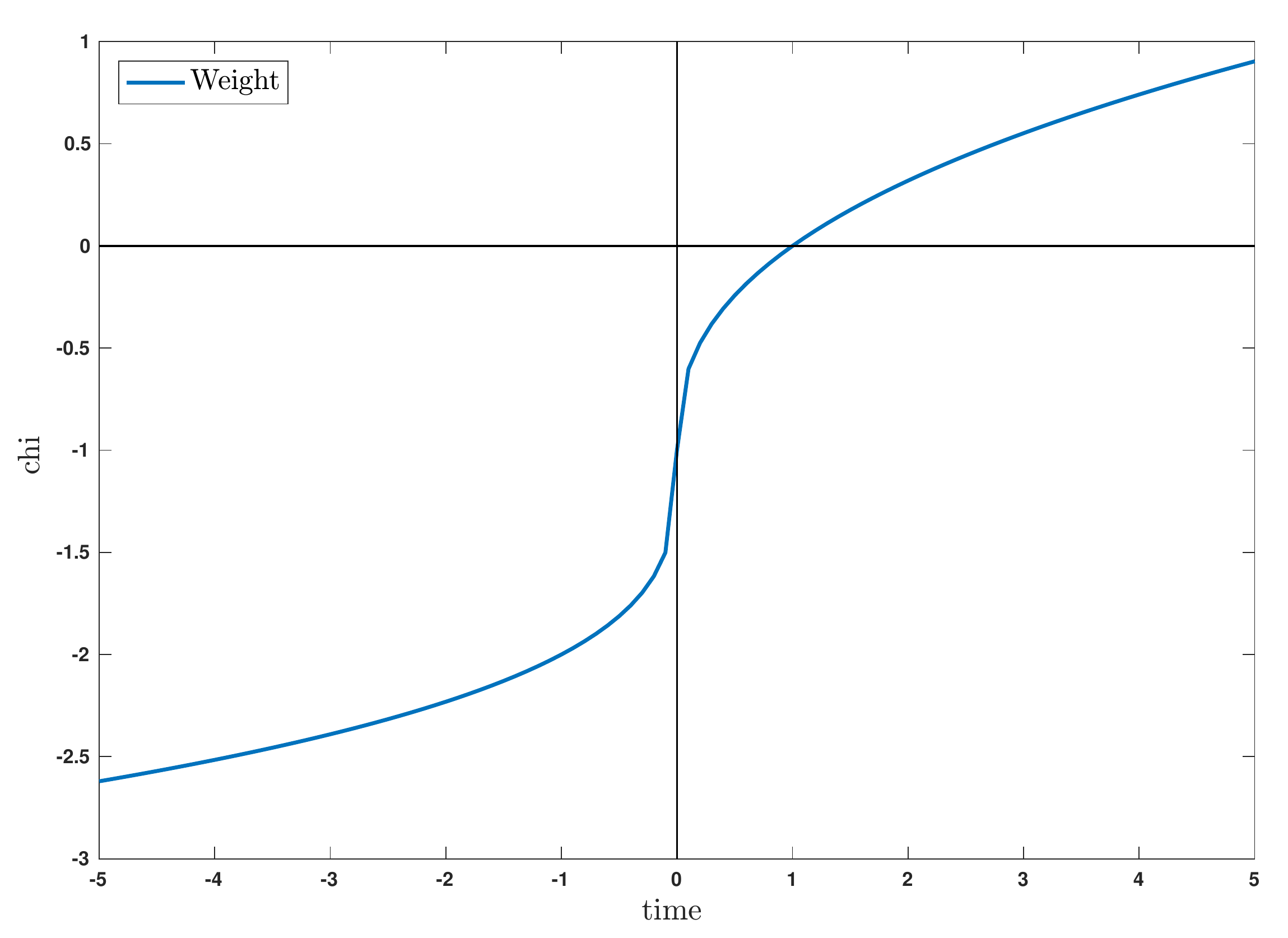}
\caption{Example of the value function $U_{\gamma}$ with different reference points.}\label{ValueFunction}
\end{center}
\end{figure}
The value function $U_{\gamma}$ models psychological behavior of an investor towards gains and losses. In order to derive the value function, one needs to measure the attitude of an individual as a deviation from a reference point. We have chosen the reference point to be zero, since $U_{\gamma}(0)=0$ holds. In Figure \ref{ValueFunction} we have plotted $U_{\gamma}$ and $\bar{U}_{\gamma}:= U_{\gamma}-1$ . The value function $\bar{U}_{\gamma}$ is an example of a value function with a negative reference point. Our choice of value function satisfies the usual assumptions: the function is concave for gains and convex for losses, which corresponds to risk aversion and risk seeking behavior of investors. Furthermore, our value function is steeper for losses than for gains, which models the psychological loss aversion of financial agents (see Figure \ref{ValueFunction} ).\\
First, we have a look at the price and portfolio dynamics in the case of long-term investors. Secondly, we consider the case of high-frequency trader. Detailed information of the parameter choices can be found in the Appendix \ref{paramMC}. \revised{We aim to give a short discussion of the model parameters. The different parameter choices in $\omega, \gamma$ and $\kappa$ for the computation of the marginals are selected in order to ensure the correct sign of the aggregated estimate of stock return over bond return $K$, which is essential in order to ensure a closed PDE for the marginals. For details we refer to section \ref{MargDiff}. The larger time step and the smaller sample size in the high-frequency case is chosen because of the additional complexity caused by the high-frequency stock price PDE.}

\begin{figure}[htb]
\begin{center}
\includegraphics[scale=0.32]{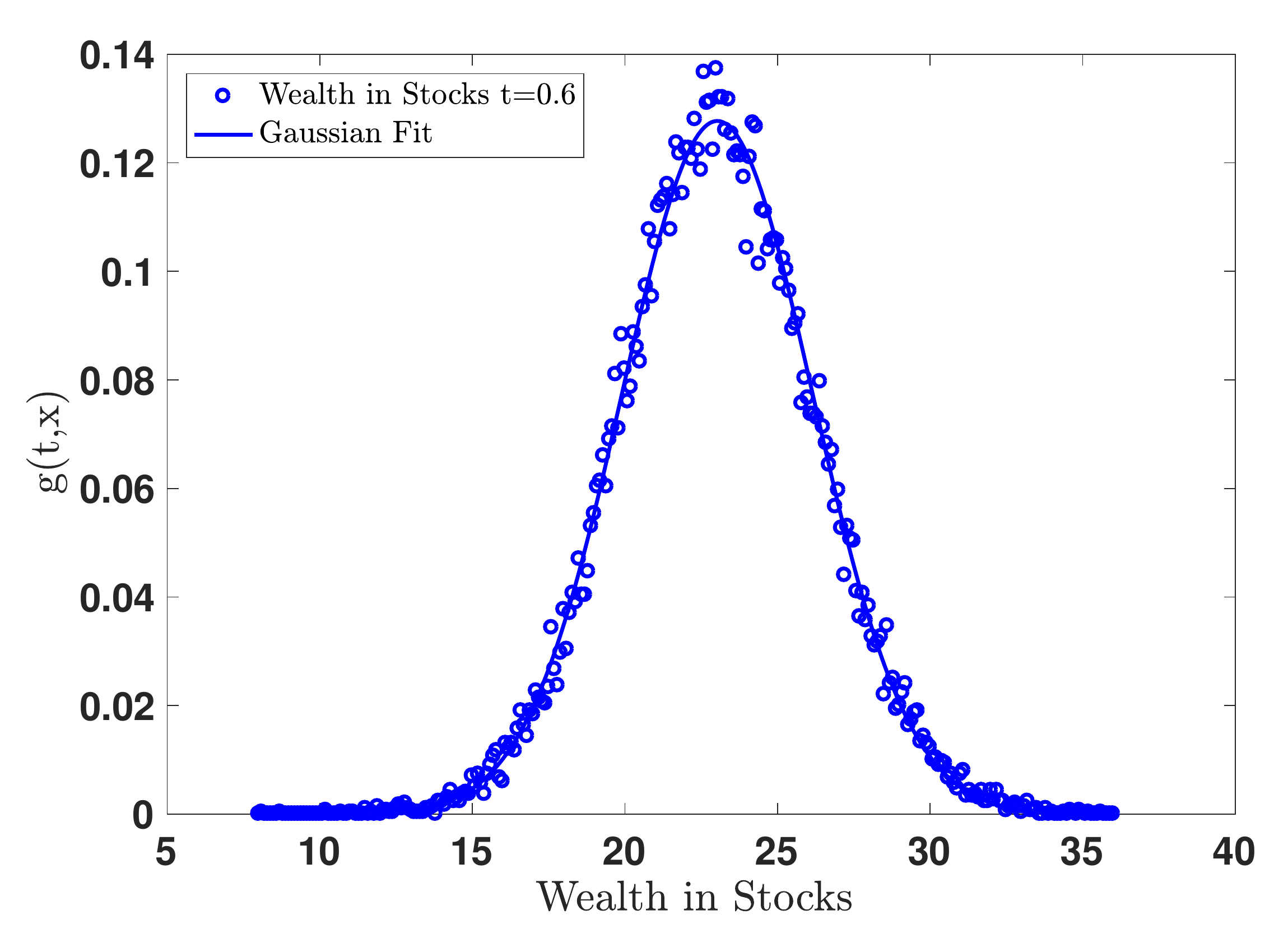} 
\hfill
\includegraphics[scale=0.32]{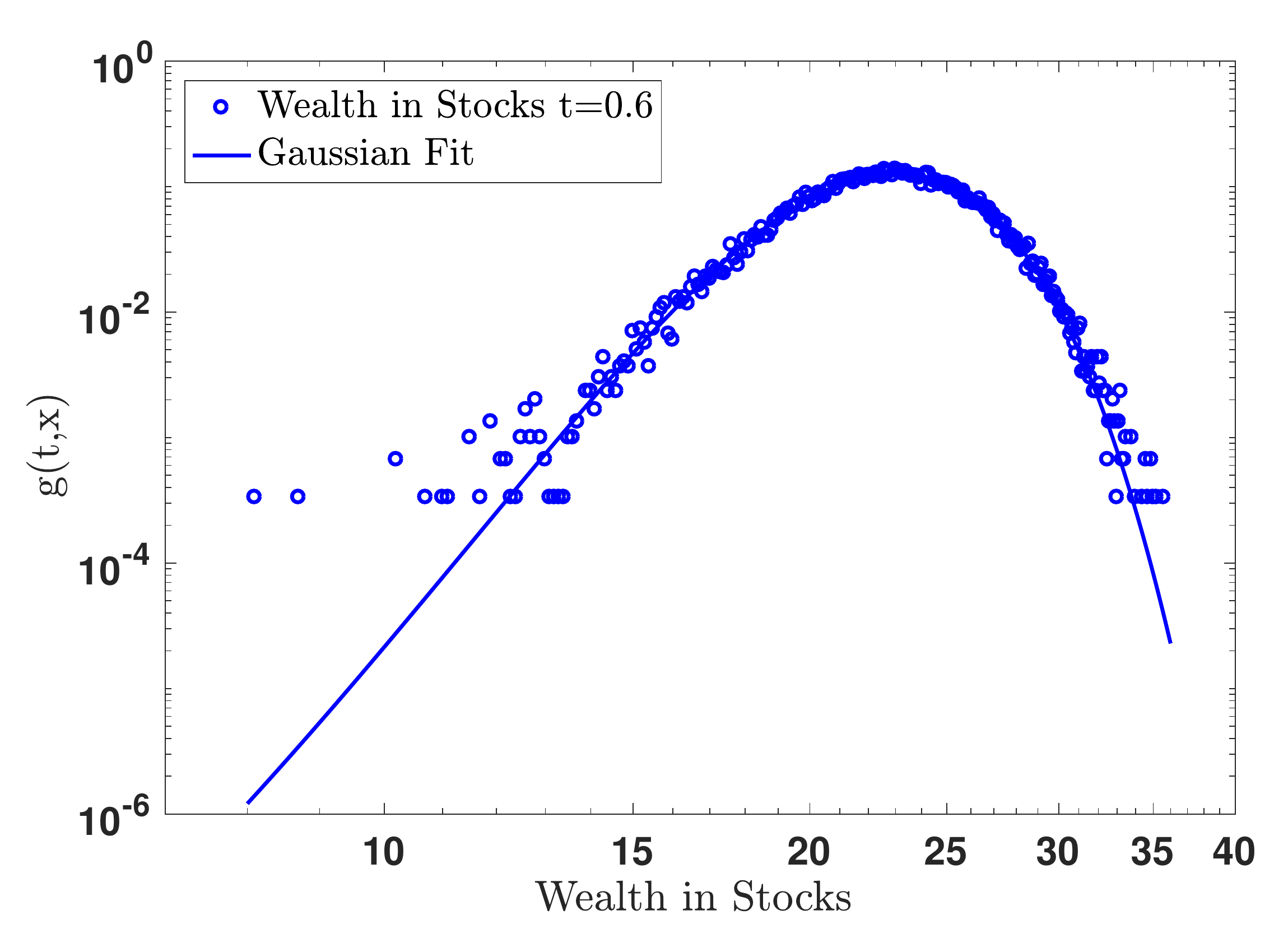} 
\caption{Distribution of the wealth invested in stocks with a Gaussian fit (solid line). Left figure has a linear scale, whereas the right figure shows the distribution in log-log scale. }\label{MargWealth}
\end{center}
\end{figure}

\begin{figure}[htb]
\begin{center}
\includegraphics[scale=0.32]{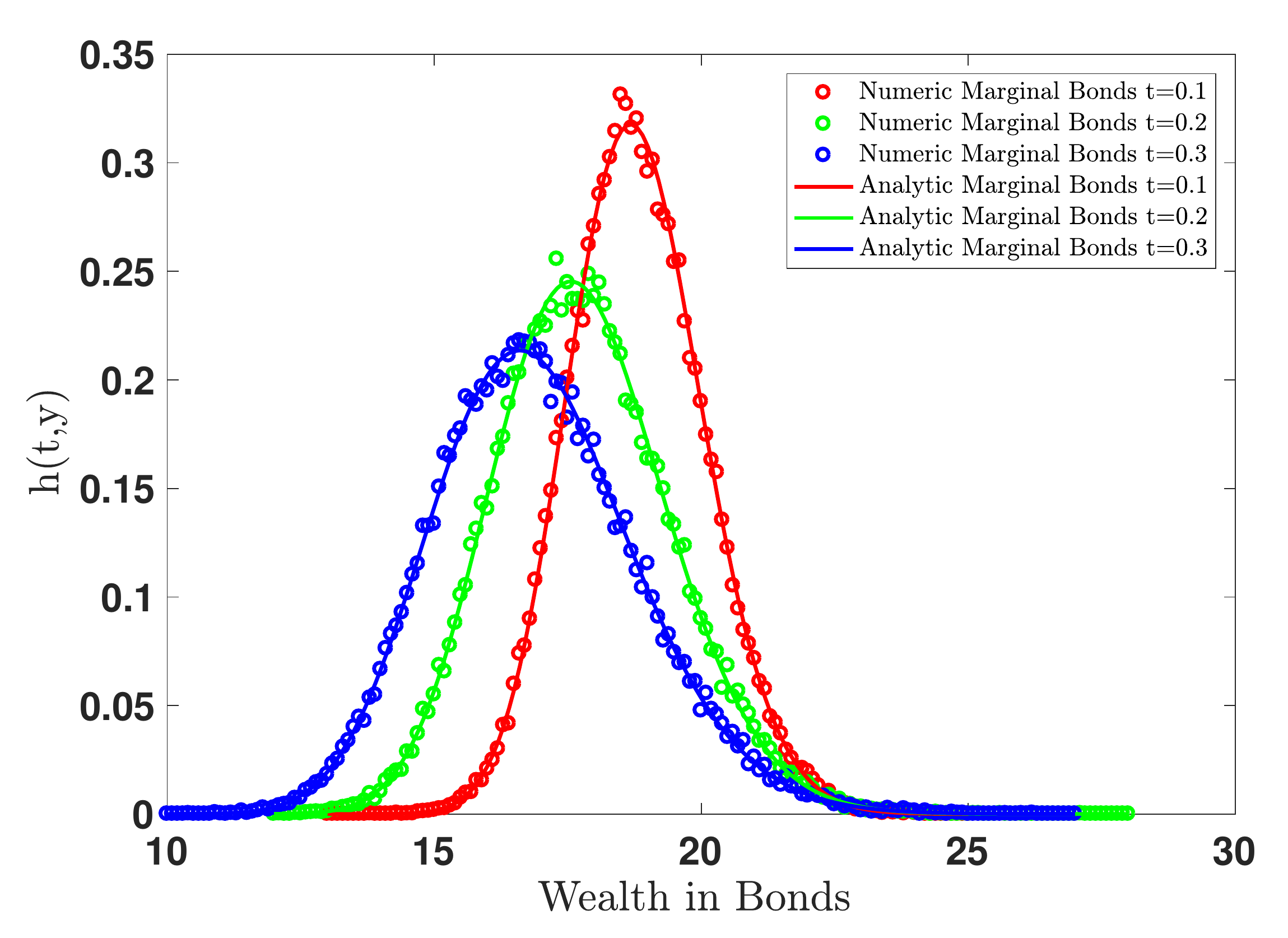} 
\hfill
\includegraphics[scale=0.32]{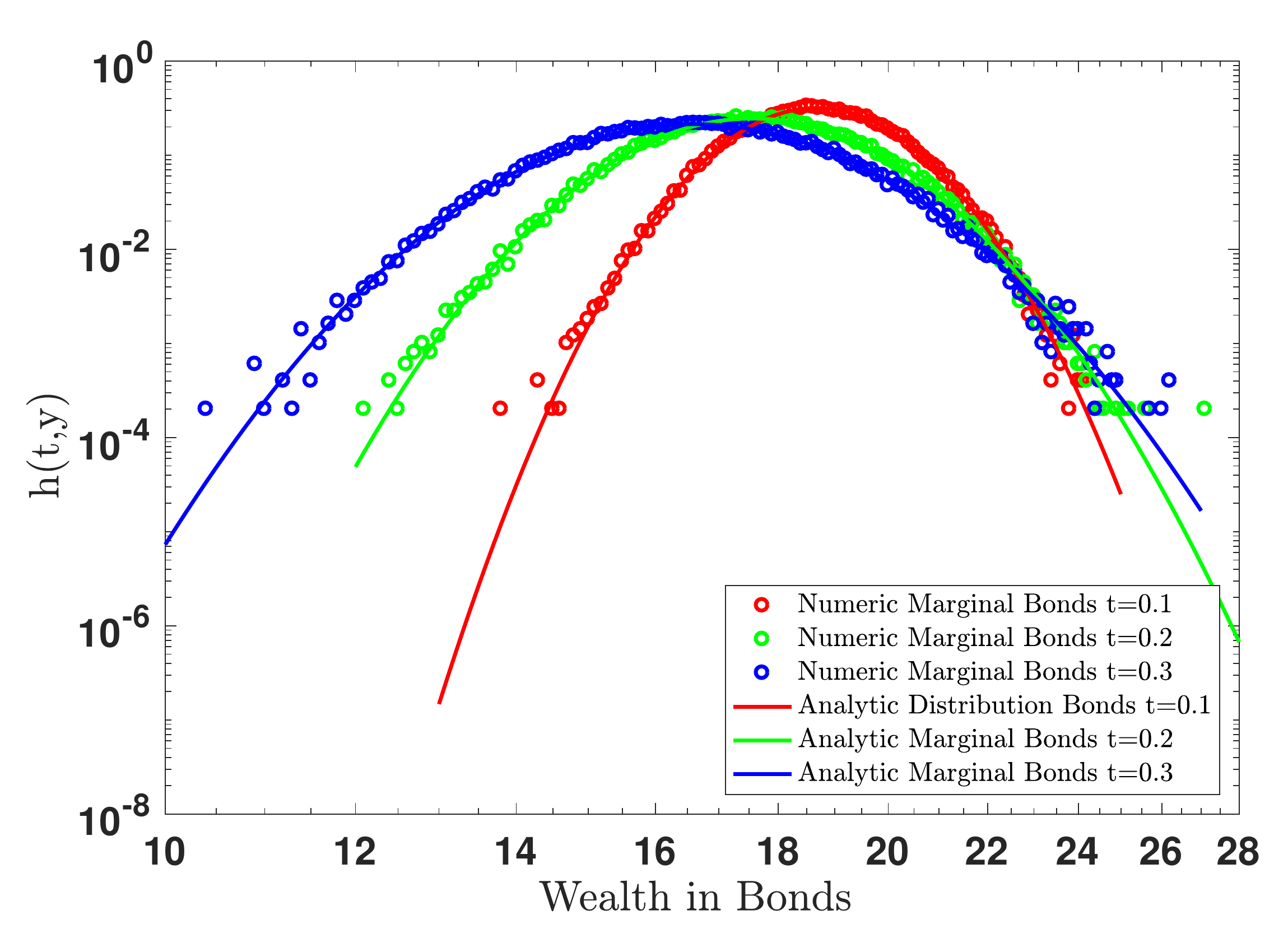} 
\caption{Distribution of the wealth invested in bonds in the special case $K>0$. The numerical results (circles) are plotted with the corresponding log-normal analytic self-similar solution (solid lines). }\label{LogAnal}
\end{center}
\end{figure}

\paragraph{Long-Term Investors.}
In order to observe more realistic price behavior, we introduce a time varying fundamental price $s^f(t)$. We choose a stationary log-normally distributed fundamental price, modeled by the following SDE
\[
ds^f = 0.1\ s^f\ dW.
\]
Again, $W$ denotes the Wiener process and the integrals need to be interpreted in the It$\hat{\text{o}}$ sense.
\begin{figure}[t]
\begin{center}
\includegraphics[scale=0.32]{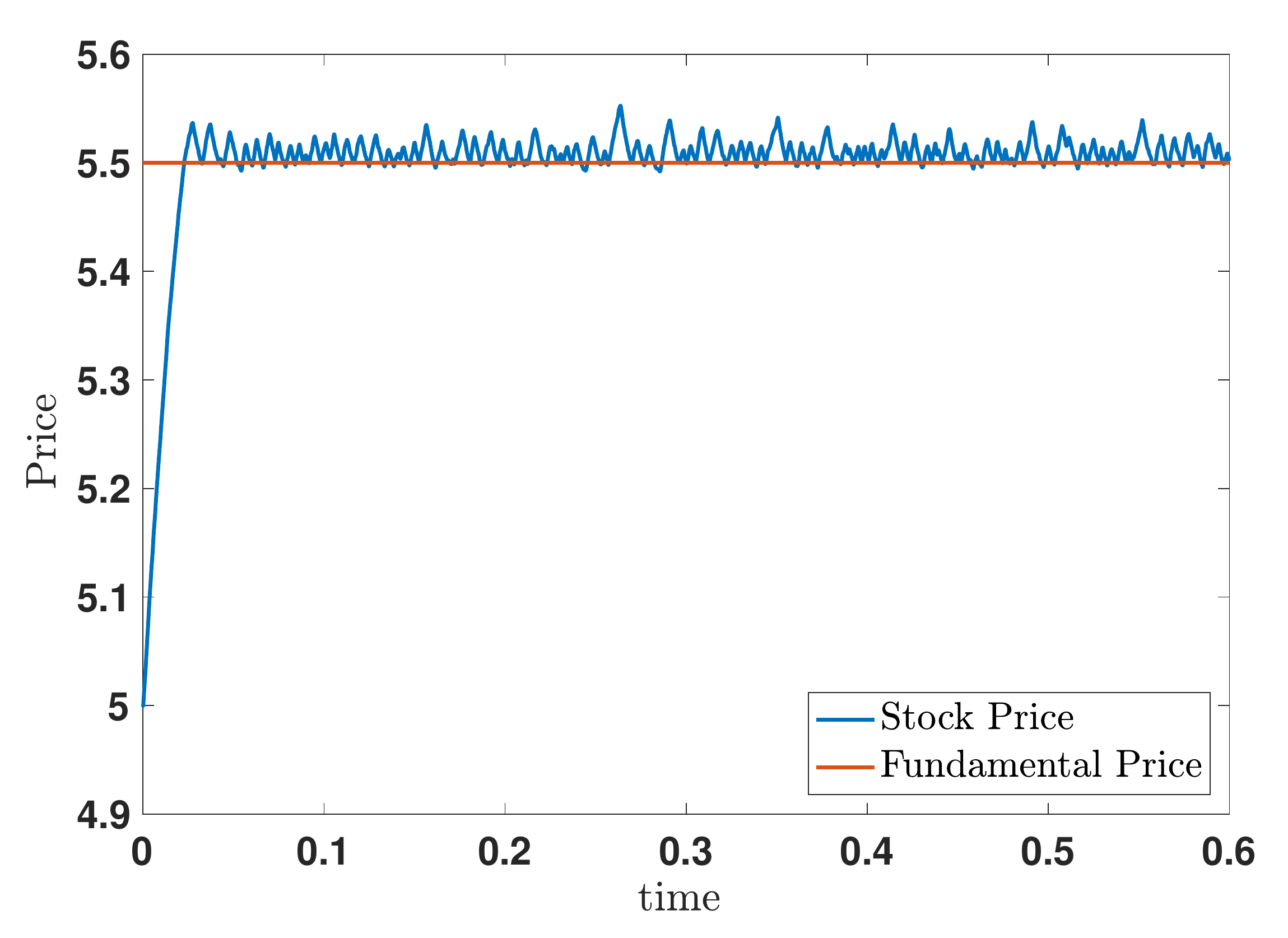} 
\hfill
\includegraphics[scale=0.32]{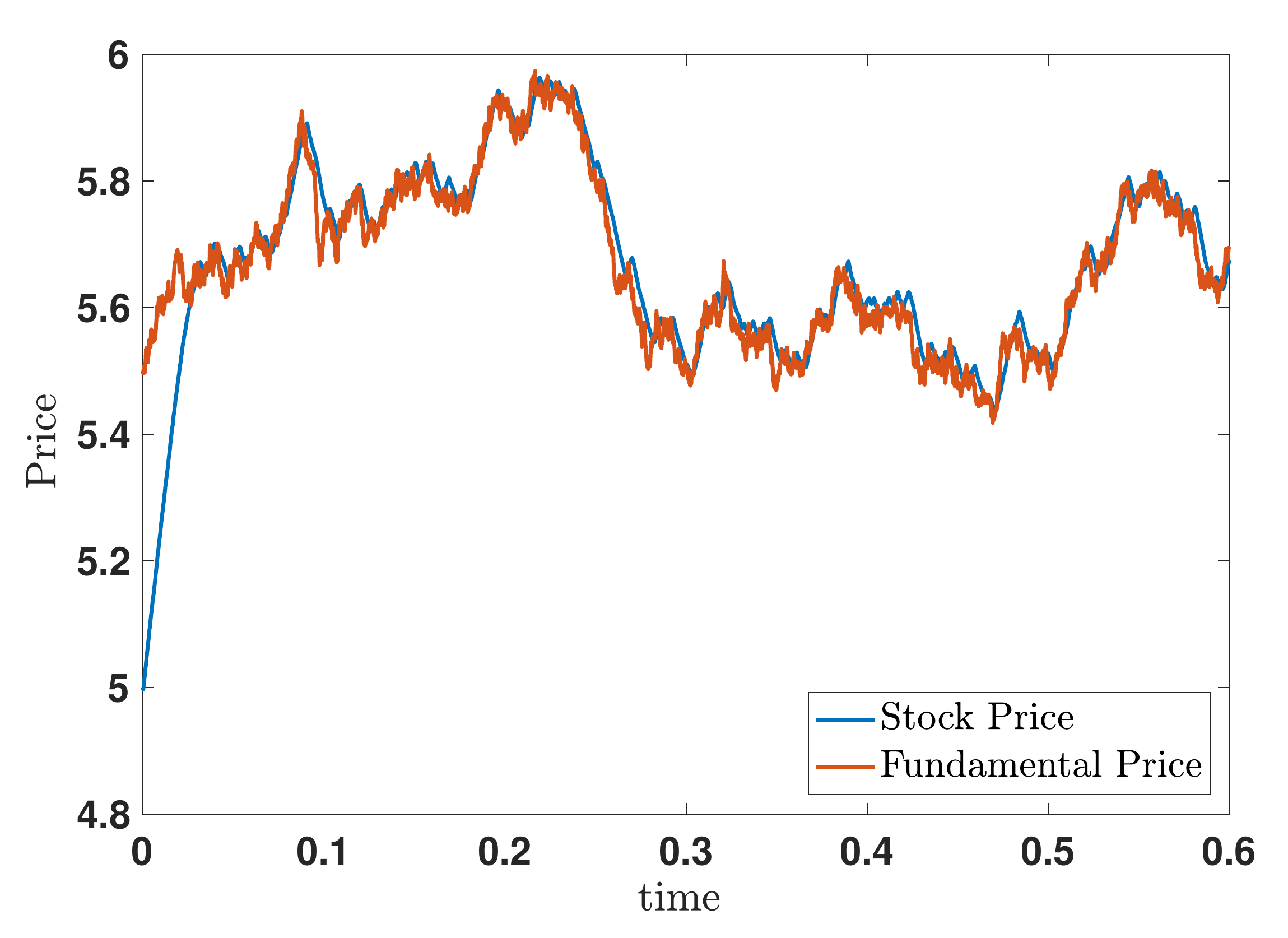} 
\end{center}
\caption{Stock price evolution in the long-term investor case with a constant fundamental price $s^f$ (left figure) and a time varying fundamental price (right figure). In both figures one obtains that the average stock price is above the funcamental value. }\label{stockFull}
\end{figure}
\begin{figure}[t]
\begin{center}
\includegraphics[scale=0.32]{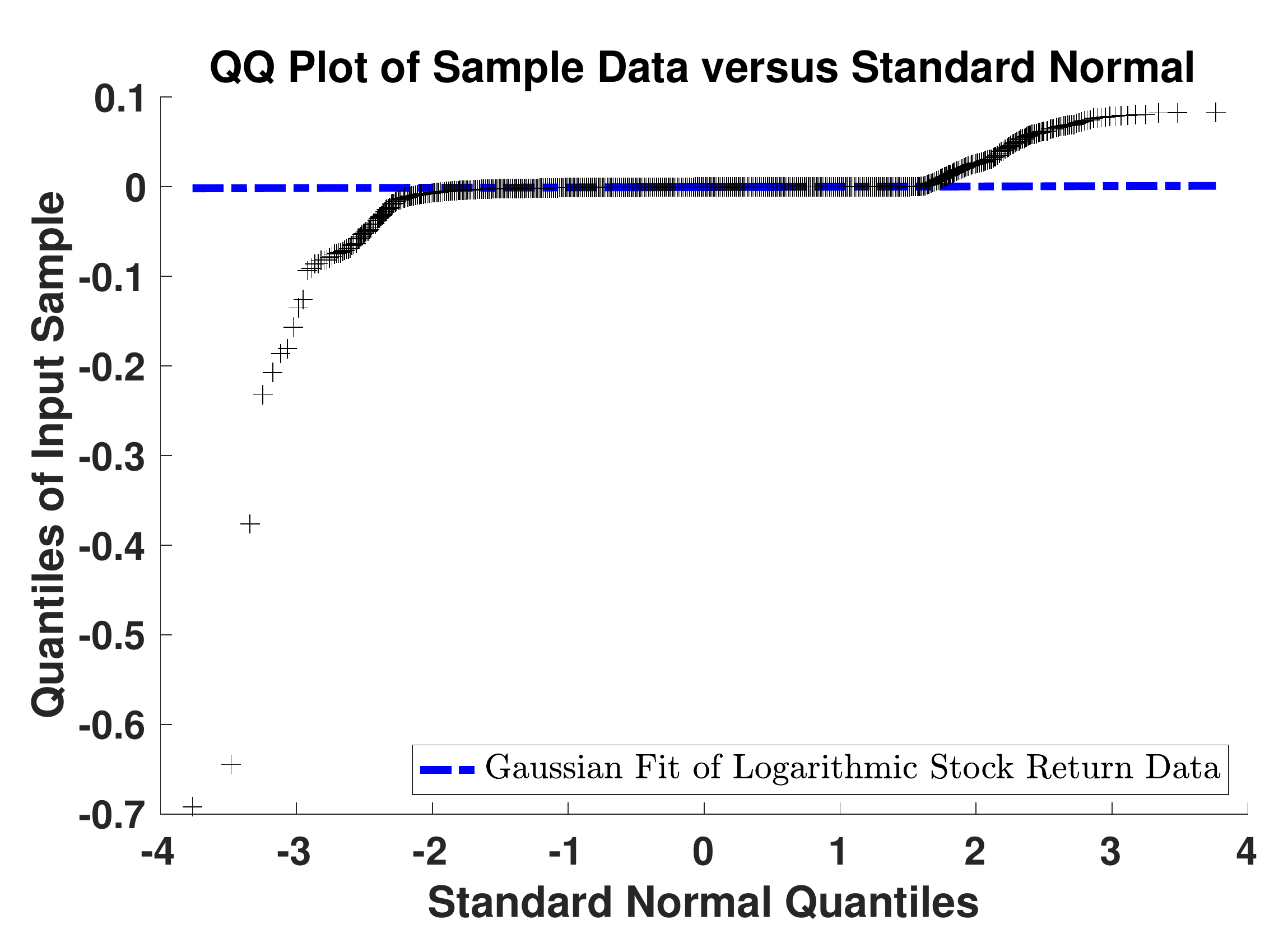} 
\hfill
\includegraphics[scale=0.32]{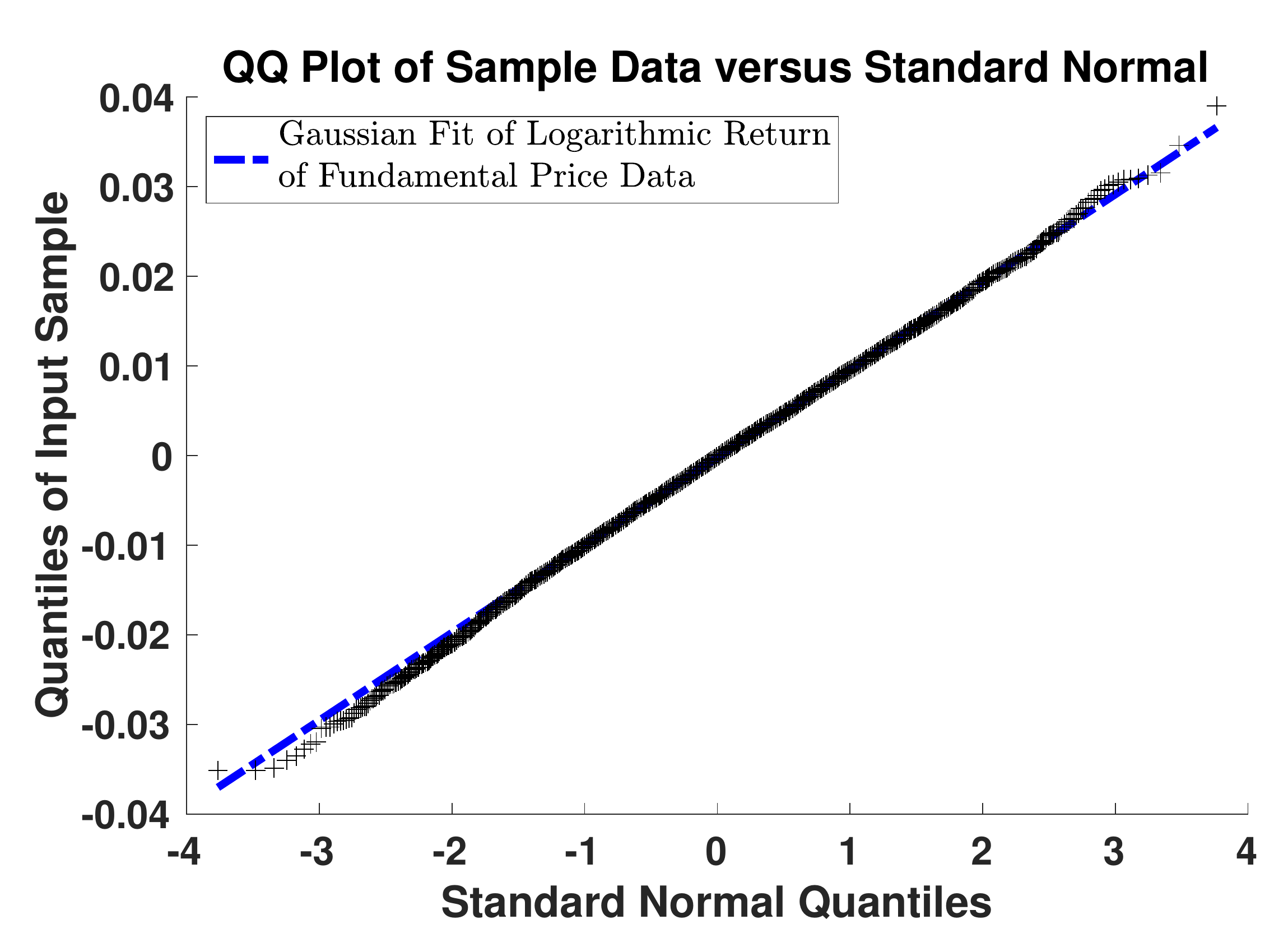} 
\caption{Quantile-quantile plot of logarithmic stock return distribution (left-hand side) and logarithmic return of fundamental prices (right-hand side). The simulation has been performed in the case of long-term investors and a stochastic fundamental price. The risk tolerance has been set to $\gamma=0.9$, the scale to $\rho=\frac{5}{8}$ and the random seed is chosen to be \texttt{rng(767)}. All further parameters are chosen as reported in section \ref{paramMC} of the Appendix.}\label{qqplot}
\end{center}
\end{figure}
In the case of a constant fundamental price, we observe oscillatory behavior (see Figure \ref{stockFull}). 
The price behavior rapidly changes for a time varying fundamental price. The stock price follows the fundamental price but has some overshoots (see Figure \ref{stockFull}). These overshoots are essential to observe a fat-tail in the stock return distribution. To quantify this, we look at a quantile-quantile plot of logarithmic stock returns. The quantile-quantile plot fits the data to the quantile of a Gaussian distributed random variable. For that reason, data which is well fitted by a Gaussian appears linear and is located on the dotted line.
We easily recognize that the stock return exhibits heavy tails (see Figure \ref{qqplot}). In comparison to the stock return, the return of fundamental prices is well fitted by a Gaussian distribution. \\
Due to the mesoscopic kinetic model, we can analyze the price and wealth distributions. In the previous paragraph, we could show that the stock price distribution is given by log-normal law. 
\begin{figure}[htb]
\begin{center}
\includegraphics[scale=0.35]{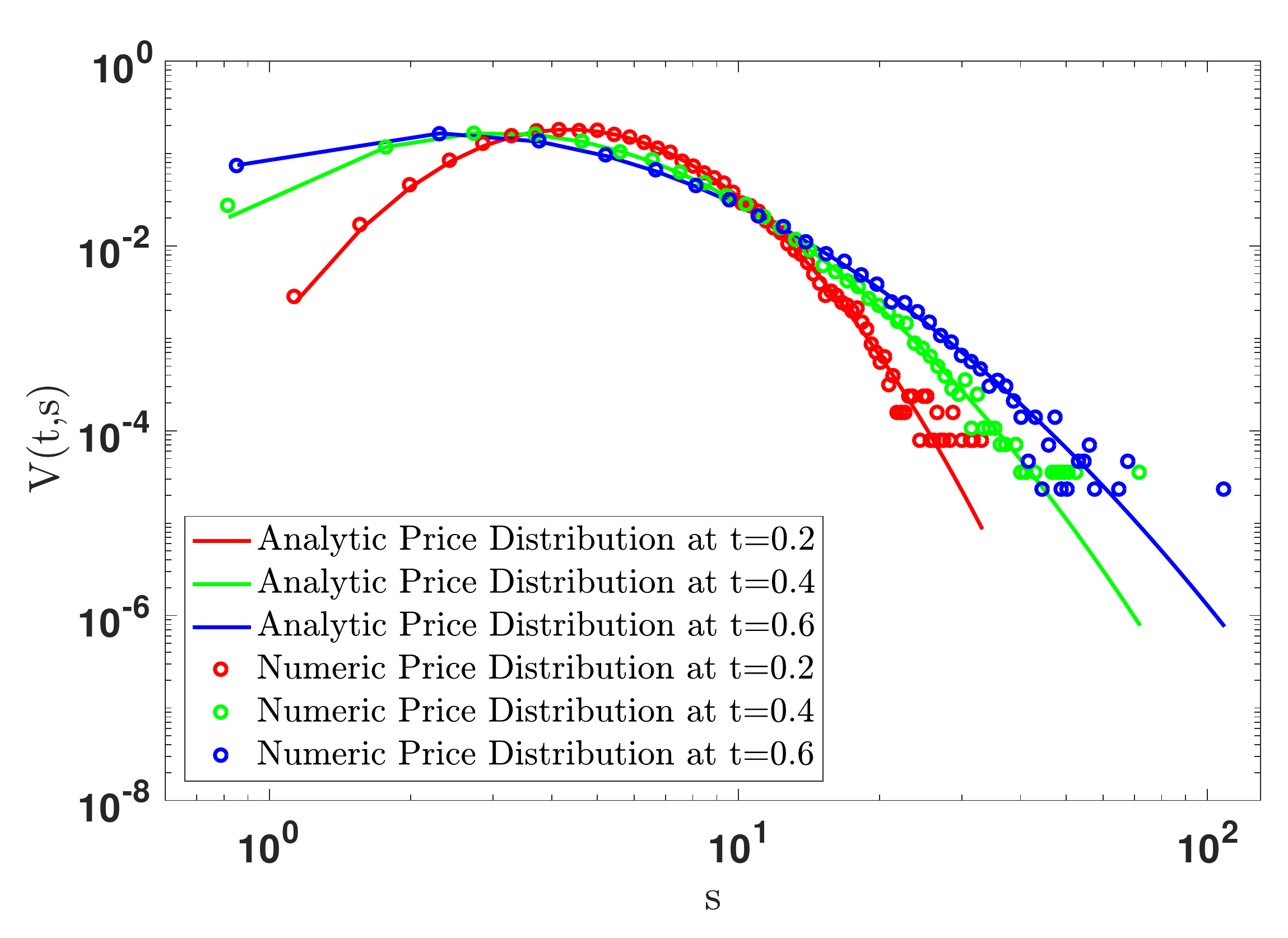} 
\caption{Stock price distribution in the long-term investor case. The solid lines are analytical solution, whereas the circles are the numerical result. }\label{lognorm}
\end{center}
\end{figure}
Our simulations (see Figure \ref{lognorm} ) verify this result. In addition, we want to have a look at the marginal distribution $g(t,x)$ which describes the wealth of the stock portfolio. 
Interestingly, the distribution of stock investments is well fitted by a normal distribution (see Figure \ref{MargWealth}). 
In the special situation that the aggregated estimate of stock return over bond return, denoted by $K$, is strictly positive or strictly negative, we can compute marginal distributions analytically. Then the marginal distribution admits log-normal behavior. For our example, we consider the case $K>0$, thus, we observe the marginal distribution of wealth in bonds $h$. In order to ensure $K>0$, we have set the fundamental stock price to $s^f\equiv 10$ and fixed the weight $\chi\equiv 1$. As Figure \ref{LogAnal} illustrates, the numerical simulations confirm the analytic results. 

\begin{figure}[htb]
\begin{center}
\includegraphics[scale=0.35]{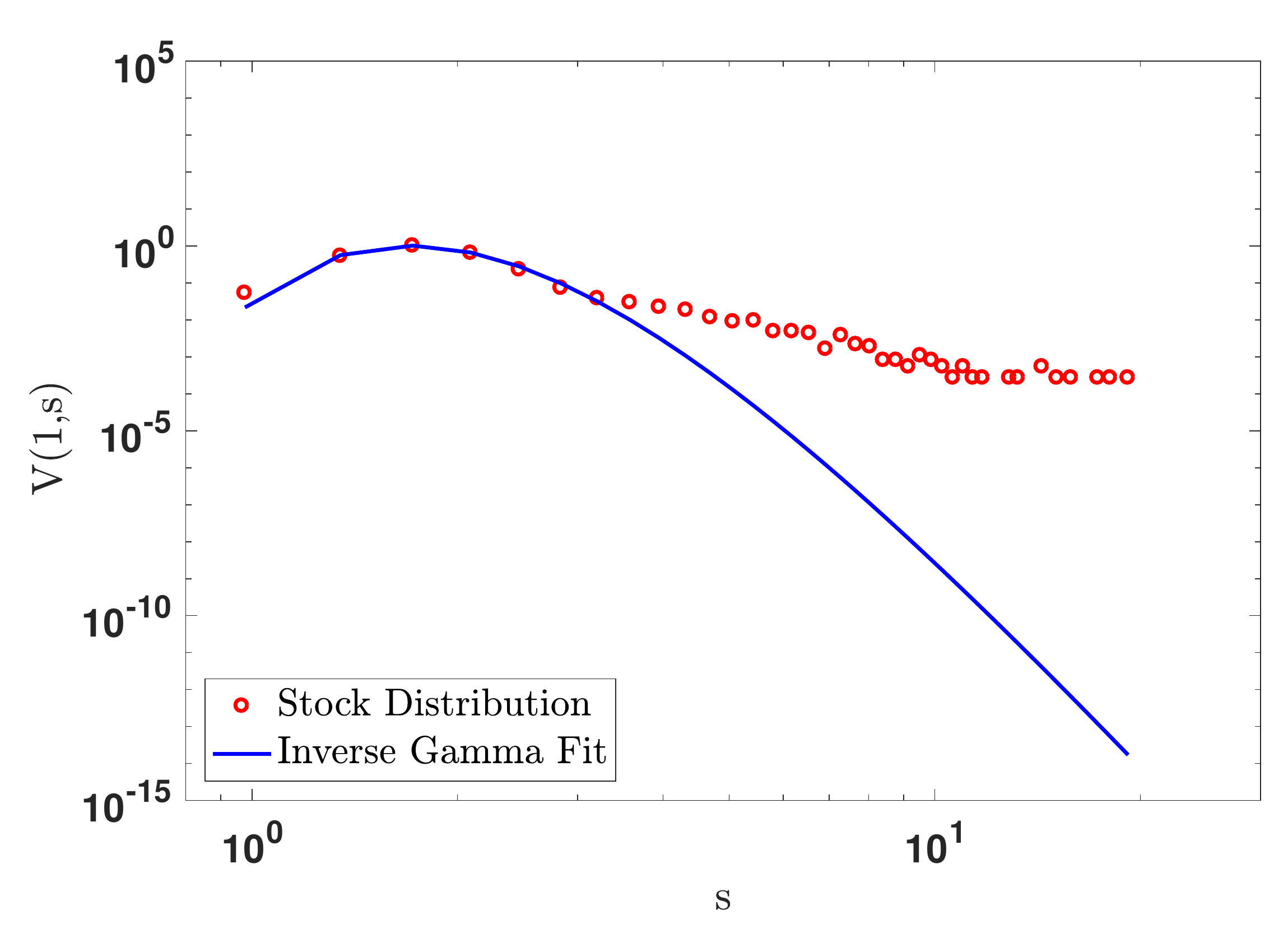} 
\caption{Stock price distribution in the high-frequency case (red circles). The fit by the inverse-gamma distribution (solid line) clearly underestimates the tail. This reveals that the full model can create heavier tails than the inverse-gamma distribution. }\label{fattailnumeric}
\end{center}
\end{figure}

\begin{figure}[htb]
\begin{center}
\includegraphics[scale=0.32]{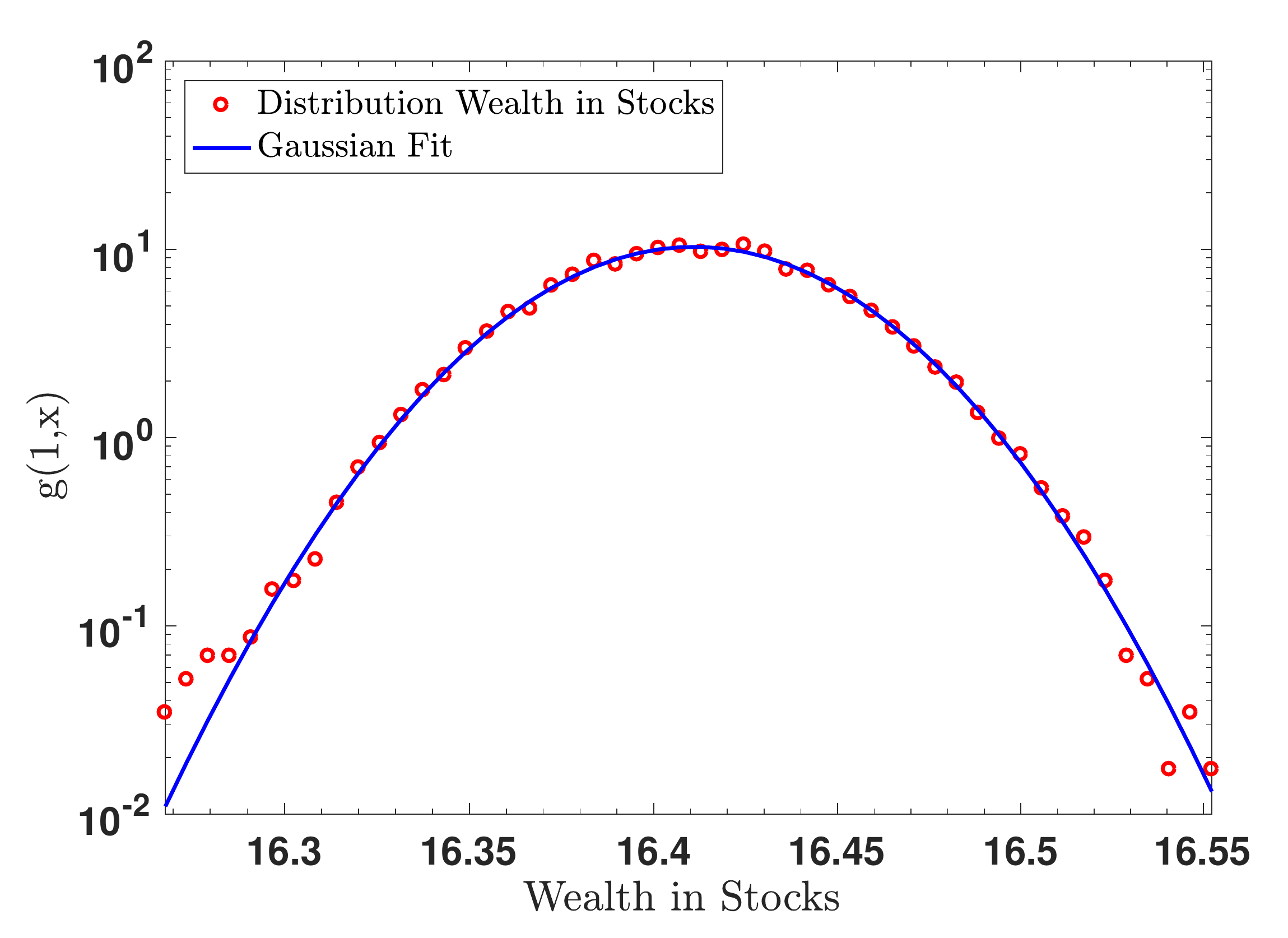} 
\hfill
\includegraphics[scale=0.32]{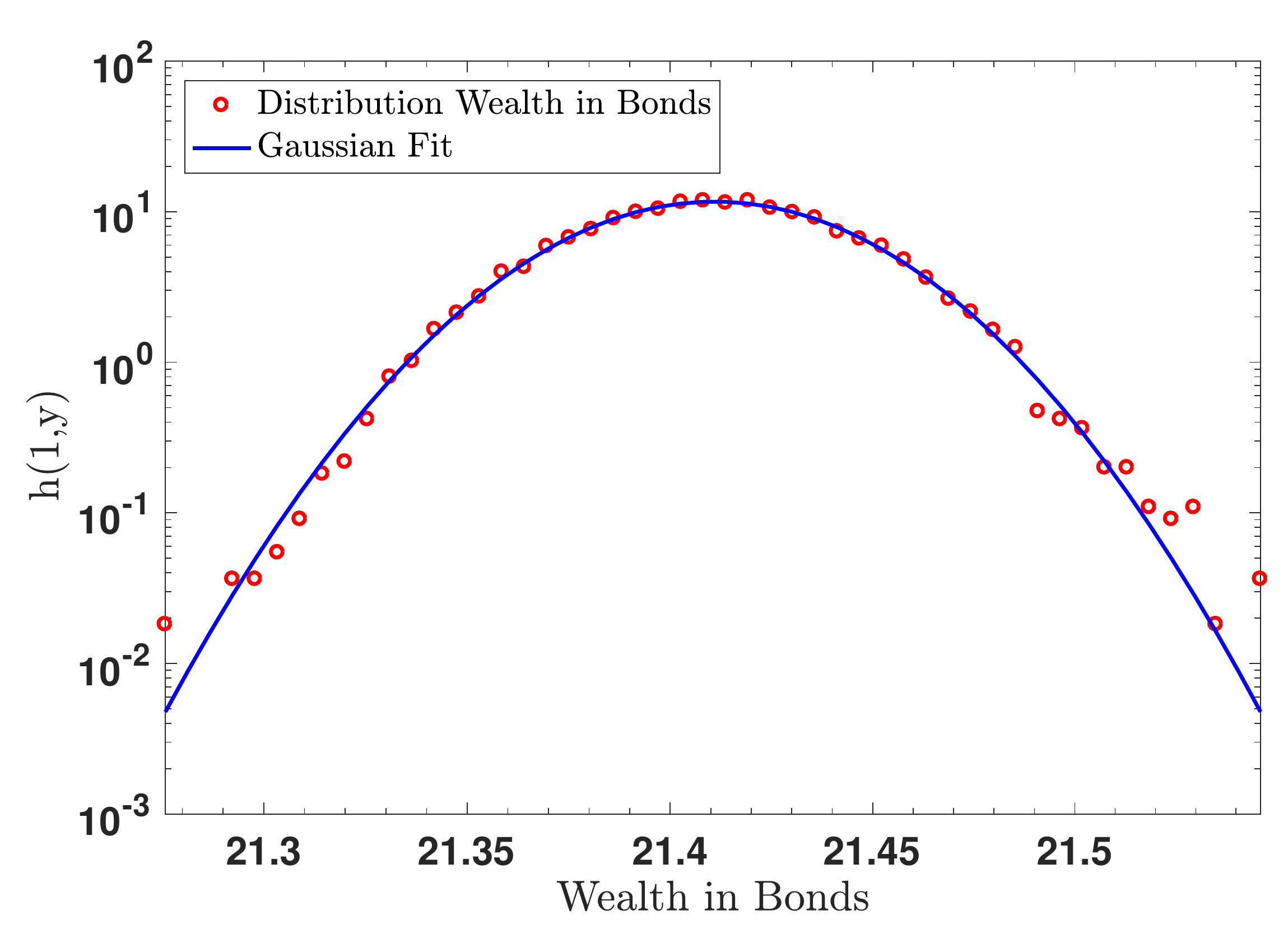} 
\caption{Marginal wealth distributions in the high-frequency investor case.  The left hand side illustrates the distribution of investments in stocks and the right-hand side the wealth invested in bonds at $t=1$. }\label{LogPortHigh}
\end{center}
\end{figure}

\paragraph{High-Frequency Investors.}
\revised{First, we study the qualitative behavior of the stock and wealth distribution. The parameters are chosen accordingly to the table in appendix \ref{paramMC}.}
In the high-frequency investor case, we numerically observe a fat-tail (see Figure \ref{fattailnumeric}). The fit by the inverse-gamma distribution reveals that the fit underestimates the tail probabilities. 
This indicates that the model can create realistic power laws. 
Furthermore, the wealth distributions are in both portfolios well-fitted by a Gaussian distribution as you can see in Figure \eqref{LogPortHigh}. The shape of the wealth coincides with the marginal portfolio distributions we computed in the long-term investor case.  

\revised{Secondly, we compute the admissible steady state of the high-frequency stock price PDE. 
In the previous section, we could compute the steady state distribution in a special case, and under the additional assumption that the portfolio dynamics have reached a steady state, analytically}.
We have observed that the inverse-gamma distribution is a steady state, which is asymptotically well characterized by a power-law for large stock prices $s$.
\revised{In order to compute the steady state numerically we have to choose several parameters different from those stated in the table in appendix \ref{paramMC}. 
More precisely, the constants $r$ and $D$ must be selected such that we obtain a steady state in the portfolio dynamics.
}
Furthermore, we do not consider the diffusive portfolio equation, but instead the mean field portfolio equation.
In addition, the value function has been chosen as the identity and the weight is fixed as $\chi\equiv 1$.
 \begin{figure}[htb]
\begin{center}
\includegraphics[scale=0.35]{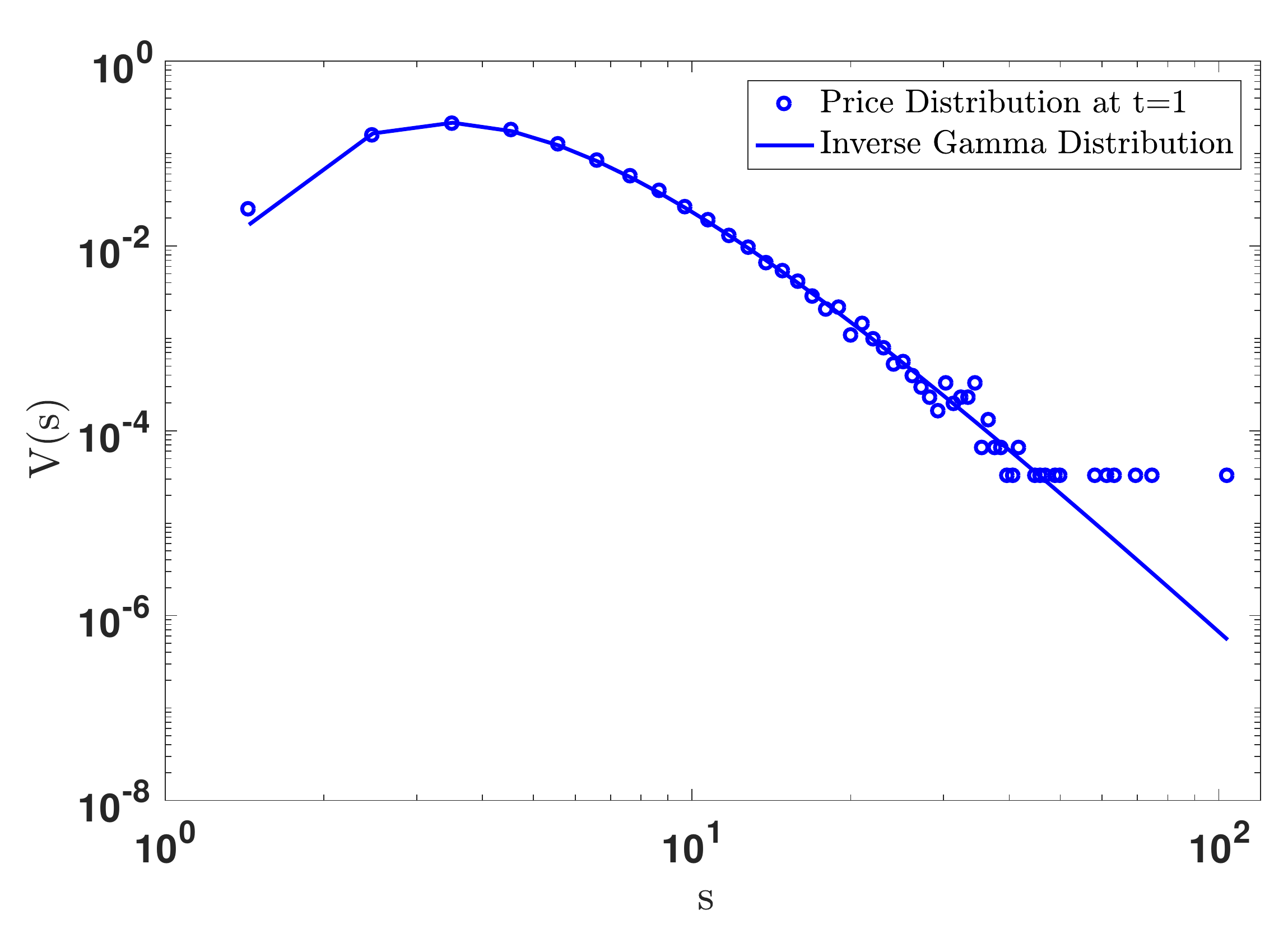} 
\caption{Steady state stock price distribution in the high-frequency investor case (circles) together with the analytically computed steady state of inverse-gamma type (solid line). }\label{invGammaNumeric}
\end{center}
\end{figure}
Figure \ref{invGammaNumeric} shows that the stock price distribution converges to the analytically computed steady state of inverse-gamma type.

\section{Conclusions}
The starting point of our investigation was a microscopic portfolio model coupled with
the macroscopic stock price equation. Each financial agent was equipped with an optimization problem in order to derive his/her investment decision. \revised{Thanks to the MPC approach applied in a game theoretical setting we simplify the optimization problem and compute the feedback control. 
Then, starting from the feedback controlled model, in the limit of a large number of agents, we derived the mean field description and the diffusive mean field description in presence of noise. 
The mean field models have been analyzed to discover insights in the portfolio distribution.} The marginal distributions of wealth in bonds or wealth in stocks 
can be characterized by a log-normal distribution in special cases. These findings have been supported by numerical simulations.
We have employed the diffusive mean field portfolio stock price model to investigate the price behavior. In the case of long-term investors, the price distribution is given by a log-normal law. In addition, we have computed a steady state of inverse-gamma type in the stock price distribution for high-frequency trader. We have seen that for large stock prices the distribution asymptotically satisfies a power-law. 
%
Interestingly, we do not observe fat-tails in the portfolio distribution.
This is quite surprising as one could expect to see the same distribution in the stock price and portfolio dynamics.
%
In order to observe a power-law in the wealth distribution, there are several possible model extensions. 
One idea is to add earnings to the microscopic model. Thus, one would add an external force on the microscopic level. This might give a fat-tail in the portfolio distribution. Alternatively, one could introduce wealth interactions among agents. There are several kinetic models which consider wealth distributions where a power-law has been observed \cite{bouchaud2000wealth, cordier2005kinetic, chatterjee2007kinetic, pareschi2006self}. \revised{Other model improvements are obtained by considering further individual features in the dynamic, like the agents' attitude towards risk or other behavioral aspects.}
We leave this question open for further research. 

\section*{Acknowledgements}
T. Trimborn acknowledges the supported by the Hans-B\"ockler-Stiftung. T. Trimborn would like to thank the
German Research Foundation DFG for the kind support within the Cluster of Excellence Internet of Production (Project-ID: 390621612).
L. Pareschi acknowledge the support of the MIUR-DAAD program \emph{Mean field games for socio economic problems}.

\appendix
\section{Appendix}

\subsection{Marginals of mean field portfolio model}\label{appMar}
For $K>0$, we get  a closed equation for $h$:
\begin{align*}
&\partial_t h(t,y)+\partial_y\left( \left( r - \frac{K(S(t))}{\nu} \right)y\  h(t,y)\right) =0. 
\end{align*}
This advection equation can also be solved by the log-normal density function
$$
h(t,y)= \frac{\hat{c}}{\sqrt{\pi} y}\ \exp\left\{-\left(\log(y)-\int\limits_0^tr - \frac{K(S(\tau))}{\nu} \ d\tau\right)^2\right\},\ \hat{c}>0, 
$$
which can be verified by simple computations. 

\subsection{Marginals of diffusive mean field portfolio model}\label{MarDiff}
In the case $K>0$, we obtain for $h$ the equation
\begin{align*}
\partial_t h(t,y)+\partial_x \left( \left[r- \frac{K(S(t))}{\nu} \right]    y\ h(t,y)\right) = \frac{1}{2\ \nu^2} \partial_y^2(y^2\ h(t,y)).
\end{align*}
Again, we consider the scaling $\bar{h}(t,\bar{y})= y\ h(t,y),\ \bar{y}=\log(y)$ and define $e(t):=r-\frac{K(S(t))}{\nu}$. Simple computations reveal that $\bar{h}$ satisfies
\begin{align*}
\partial_t \bar{h}  (t,\bar{x})+ \left(e(t)-\frac{1}{2\ \nu^2}\right) \ \partial_{\bar{y}} h(t,\bar{y}) = \frac{1}{2\ \nu^2} \partial_{\bar{y}}^2 \bar{h}(t,\bar{y}).
\end{align*}
We define $E(t):= \int\limits_0^t e(\tau)\ d\tau +c_2,\ c_2>0$ and 
$$
\bar{h}(t,\bar{y})= \frac{1}{(2\ (\frac{t}{\nu^2}+c_1)\ \pi)^{\frac12}} \exp\left\{- \frac{(\bar{y}+\frac{(\frac{t}{\nu^2}+c_1)}{2}-E(t))^2}{2\ (\frac{t}{\nu^2}+c_1)}\right\},\ c_1>0,
$$
solves the previous convection-diffusion equation. Then, reverting to the original variables, we observe a log-normal law.
$$
h(t,y)= \frac{1}{y\ (2 (\frac{t}{\nu^2}+c_1)\ \pi)^{\frac12}}  \exp\left\{ -\frac{\left(\log(y)+\frac{\frac{t}{\nu^2}+c_1}{2}-E(t)\right)^2}{2\ (\frac{t}{\nu^2}+c_1)}\right\},\ c_1>0.
$$

\subsection{Asymptotic limit of Boltzmann model} \label{asymp}
We expand the test function $\phi(x^{\prime},y^{\prime})$ in a Taylor series up to order two and we denote by $R$ the remainder of the Taylor series.
The right-hand side of the kinetic equation is then given by:
\begin{align*}
 (L(f),\phi)=& \Big\langle \theta \int a \left[ x\ \left(\kappa\ ED(t,f,S)+\frac{D(t)}{S(t)}\right) + u_{\eta}^{*}(t,x,y,S) \right] \frac{\partial \phi(x,y)}{\partial x}f(t,x,y)\  dx dy \Big\rangle +\\
& \left\langle \theta \int a\ [ y\ r- u_{\eta}^*(t,x,y,S)]\ \frac{\partial \phi(x,y)}{\partial y} f(t,x,y)\  dx dy \right\rangle +\\
& \left\langle \theta \int \left[   (x^{\prime}-x)\ (y^{\prime}-y)\ \frac{\partial^2 \phi(x,y)}{\partial y \partial x} + \frac12 (y^{\prime}-y)^2\ \frac{\partial^2 \phi(x,y)}{\partial y ^2 }\right]\  f(t,x,y)\  dx dy \right\rangle +\\
& \left\langle \theta \int \left[ \frac12 (x^{\prime}-x)^2\ \frac{\partial^2 \phi(x,y)}{\partial x^2 }+R(t,x,y)\right]\  f(t,x,y)\  dx dy \right\rangle.
\end{align*}

We make the following scaling assumptions:
\begin{align*}
\theta= \frac{1}{\epsilon},\quad a= \epsilon.
\end{align*}
The interaction operator is consequently given by:
\begin{align*}
 (L(f),\phi)=&    \int \left[  x\  \left( \kappa\ ED(t,f,S)+\frac{{D(t)}}{S(t)}\right)+ u^*(t,x,y,S)\right]\ \frac{\partial \phi(x,y)}{\partial x} f(t,x,y)\  dx dy \\
& +\int [  y\ r- u^*(t,x,y,S)]\ \frac{\partial \phi(x,y)}{\partial y} f(t,x,y)\  dx dy \\
&  +\int \epsilon \left[ \ x\  \left( \kappa\ ED(t,f,S)+\frac{{D(t)}}{S(t)}\right)\  y\ r\right]\ \frac{\partial^2 \phi(x,y)}{\partial y \partial x} \  f(t,x,y)\  dx dy\\
&  +\int \epsilon\ u^*(t,x,y,S)\ (- x)\  \left( \kappa\ ED(t,f,S)+\frac{{D(t)}}{S(t)}\right)\ \frac{\partial^2 \phi(x,y)}{\partial y \partial x} \  f(t,x,y)\  dx dy\\
&  +\int \epsilon\ u^*(t,x,y,S)\ \left[r\ y - u^*(t,x,y,S)\right]\ \frac{\partial^2 \phi(x,y)}{\partial y \partial x} \  f(t,x,y)\  dx dy\\
& -\int \frac{1}{\nu^2}(H(-K)x+H(K)y)^2 \ \frac{\partial^2 \phi(x,y)}{\partial y \partial x} \  f(t,x,y)\  dx dy  \\
&  +\int \frac{\epsilon}{2} [ r\ y+ u^*(t,x,y,S)]^2\ \frac{\partial^2 \phi(x,y)}{\partial y^2 }\  f(t,x,y)\  dx dy \\
&  +\int  \frac{1}{2\ \nu^2}(H(-K)x+H(K)y)^2 \ \frac{\partial^2 \phi(x,y)}{\partial y ^2 }\  f(t,x,y)\  dx dy \\
& + \int \frac{\epsilon}{2}  \left( x \left( \kappa\ ED(t,f,S)+\frac{D(t)}{S(t)}\right) +u^*(x,y,S)\right)^2 \frac{\partial^2 \phi(x,y)}{\partial x^2 }  f(t,x,y)\  dx dy\\
&+ \int  \frac{1}{2\ \nu^2}(H(-K)x+H(K)y)^2 \frac{\partial^2 \phi(x,y)}{\partial x^2 }  f(t,x,y)\  dx dy\\
& + R_{\epsilon}(t).
\end{align*}

Here, we have used the fact that the random variable has zero mean.
We assume that the remainder 
\[
R_{\epsilon}(t):= \left\langle  \int R_{\epsilon}(x,y)\  f(t,x,y)\  dx dy \right\rangle,
\]
vanishes in the limit $\epsilon \rightarrow 0$. Consequently, the integral operator simplifies to

\begin{align*}
 (L(f),\phi)=&   \int \left[  x\  \left( \kappa\ ED(t,f,S)+\frac{{D(t)}}{S(t)}\right)+ u^*(t,x,y,S)\right]\ \frac{\partial \phi(x,y)}{\partial x}f(t,x,y)\  dx dy  +\\
& \int [  y\ r- u^*(t,x,y,S)]\ \frac{\partial \phi(x,y)}{\partial y} f(t,x,y)\  dx dy +\\
&  \int \frac12 \frac{1}{\nu^2}(H(-K)x+H(K)y)^2 \left[ \frac{\partial^2 \phi(x,y)}{\partial x^2} +\frac{\partial^2 \phi(x,y)}{\partial y^2}-\frac{\partial^2 \phi(x,y)}{\partial y \partial x}\right]  f(t,x,y)\  dx dy,
\end{align*}

as $\epsilon \rightarrow 0$. Then, integration by parts leads to the weak form  of the following Fokker-Planck equation

\begin{align*}
&\frac{\partial}{\partial t} f(t,x,y) +\frac{\partial}{\partial x} \left(  \left[  x\  \left( \kappa\ ED(t,f,S)+\frac{{D(t)}}{S(t)}\right)+ u^*(t,x,y,S)\right]\ f(t,x,y) \right) +\\ &\frac{\partial}{\partial y} \left(  [  y\ r- u^*(t,x,y,S)] \ f(t,x,y)  \right)+ \frac{1}{\nu^2\ 2} \frac{\partial^2}{\partial x \partial y} ((H(-K)x+H(K)y)^2\  f(t,x,y))\\
 =& \frac{ 1}{\nu^2\ 2} \frac{\partial^2}{\partial x^2} ((H(-K)x+H(K)y)^2\  f(t,x,y)) +  \frac{ 1}{\nu^2\ 2} \frac{\partial^2}{\partial y^2} ((H(-K)x+H(K)y)^2\  f(t,x,y)).
\end{align*}

\subsection{Simulation parameters}\label{paramMC}
\begin{table}[h!]
\begin{subtable}{0.3\linewidth}
\small
\begin{tabular}{|c|c||c|c|}
\hline
$\Delta t$ & $0.0001$ & $\kappa$ & $0.4$ \\
\hline
$D$ &  $0.01$ & $\nu$ &$ 5$ \\
\hline
$r$ &  $0.01$ & $S_0$ & $ 5$ \\
\hline
$\alpha $ &  $0.5$ & $Y_0$ & $ 20$ \\
\hline
 $\beta$& $0.65$   & $X_0$&  $20$ \\
\hline
$\omega$ &  $80$  & $T_{end}$ & $0.6$   \\
\hline
$\gamma$  & $0.55$  & N &  $3\cdot 10^4$  \\
\hline
$\rho$  & $\frac{2}{3}$  & &     \\
\hline
\end{tabular} 
\caption*{Random Fundamental Price.}
\end{subtable}
\hfill
\begin{subtable}{0.3\linewidth}
\small
\begin{tabular}{|c|c||c|c|}
\hline
$\Delta t$ & $0.0001$ & $\kappa$ & $0.1$ \\
\hline
$D$ &  $0.01$ & $\nu$ &$ 5$ \\
\hline
$r$ &  $0.01$ & $S_0$ & $ 5$ \\
\hline
$\chi $ &  $1$ & $Y_0$ & $ 20$ \\
\hline
 $s^f$& $10 $   & $X_0$&  $20$ \\
\hline
$\omega$ &  $20$  & $T_{end}$ & $0.3$   \\
\hline
$\gamma$  & $0.35$  & N & $5\cdot 10^4$    \\
\hline
$\rho$  & $\frac{2}{3}$  & &     \\
\hline
\end{tabular} 
\caption*{Computation of Marginal.}
\end{subtable}
\hfill
\begin{subtable}{0.3\linewidth}
\small
\begin{tabular}{|c|c||c|c|}
\hline
$\Delta t$ & $0.001$ & $\kappa$ & $0.4$ \\
\hline
$D$ &  $0.01$ & $\nu$ &$ 5$ \\
\hline
$r$ &  $0.01$ & $S_0$ & $ 5$ \\
\hline
$\alpha $ &  $1$ & $Y_0$ & $ 20$ \\
\hline
 $\beta$& $0.2$   & $X_0$&  $20$ \\
\hline
$\omega$ &  $80$  & $T_{end}$ & 1   \\
\hline
$\gamma$  & $0.55$  &$N$ &  $5*10^3$  \\
\hline
$\rho$  & $\frac{2}{3}$  &$M$ &  $3*10^4$     \\
\hline
\end{tabular} 
\caption*{High-frequency trader.}
\end{subtable}
\end{table}

\newpage

	\bibliographystyle{abbrv}	
	{\small\bibliography{MPC-LLS_rev3.bib}}

\end{document}